\documentclass[journals]{IEEEtran}
\makeatletter
\def\ps@headings{%
\def\@oddhead{\mbox{}\scriptsize\rightmark \hfil \thepage}%
\def\@evenhead{\scriptsize\thepage \hfil \leftmark\mbox{}}%
\def\@oddfoot{}%
\def\@evenfoot{}}
\makeatother
\usepackage{graphicx}
\usepackage{amssymb,amsmath,latexsym,amsfonts}
\usepackage{epsfig}
\usepackage{comment}
\usepackage{color}
\newtheorem{lemma}{Lemma}
\newtheorem{conj}{Conjecture}

\newtheorem{corollary}{Corollary}
\def\done{\hspace*{\fill} \rule{1.8mm}{2.5mm}}

\begin{document}
%
\title{Flow Level QoE of Video Streaming in Wireless Networks
\thanks{Part of this work appeared in IEEE Infocom 2013.}}


\author{\authorblockN{Yuedong Xu, Salaheddine Elayoubi, Eitan Altman, Rachid El-Azouzi, Yinghao Yu}

\thanks{ Yuedong Xu and Yinghao Yu are with Department of Electronic Engineering, Fudan University, Shanghai, China. 
Rachid El-Azouzi are with LIA, Universite d'Avignon,
339 Chemin des Meinajaries, Avignon, France.
Salaheddine Elayoubi is with Orange Labs, Moulineaux, France. 
Eitan Altman is with Maestro Project-Team, INRIA Sophia Antipolis,
2004 route des Lucioles, Sophia Antipolis, France.
Email: ydxu@fudan.edu.cn, salaheddine.elayoubi@orange-ftgroup.com,
rachid.elazouzi@univ-avignon.fr, eitan.altman@inria.fr, 10300700057@fudan.edu.cn}}
\maketitle

\begin{abstract}

The Quality of Experience (QoE) of streaming service is often degraded
by frequent playback interruptions. To mitigate the interruptions, the media player prefetches streaming contents before starting playback,
at a cost of delay.
We study the QoE of streaming from the perspective
of flow dynamics. First, a framework
is developed for QoE when streaming users join the network randomly and leave after downloading completion.
We compute the distribution of prefetching delay using partial differential equations (PDEs), and the probability
generating function of playout buffer starvations using ordinary differential equations (ODEs) for CBR streaming.
Second, we extend our framework to characterize the throughput variation caused by opportunistic scheduling at the base station, and the
playback variation of VBR streaming.
Our study reveals that the flow dynamics is the fundamental reason of playback starvation.
The QoE of streaming service is dominated by the first moments such as the average throughput of opportunistic scheduling
and the mean playback rate. While the variances
of throughput and playback rate have very limited impact on starvation behavior.

\end{abstract}

\IEEEpeerreviewmaketitle

\section{Introduction}

Streaming services are witnessing a rapid growth in mobile networks. According to Allot Communications \cite{allot}, HTTP streaming service made up 37 percent of mobile broadband traffic during the second half of 2010. This presents new challenges for operators that are used to classify services into real-time (voice-like) and elastic (data-like) services. Indeed, classical QoS metrics in mobile networks are blocking rates for real-time traffic and average user throughput for elastic one, and operators dimension their networks for satisfying targets on those metrics \cite{Hegde}. However, the particular nature of streaming applications, halfway between real-time and elastic services, is raising
the following difficult questions in wireless environments. First, which QoS metrics best represent the QoE perceived by users. Second, how to predict these QoE metrics for a given traffic intensity and to dimension the network accordingly.

The first step towards defining QoE and predicting it is to understand how streaming is played. In general, media players at the devices are equipped with a playout buffer that stores arriving packets. As
long as there are packets in the buffer, the video is played smoothly. Once the buffer empties, the spacing between packets does not follow
the original one. These {\em starvations} cause large {\em jitters} and are particularly annoying for end users that see frozen images. One feasible way to avoid starvations is to introduce a start-up (also called prefetching) delay before playing the stream,
and a rebuffering delay after each starvation event. Then after a
number of media
frames accumulate in the buffer, the media player starts to work. This leads to two important sets of QoE metrics: starvation properties (probability, frequency, etc.) and startup/re-buffering delays.

Once the behavior of media streaming service is understood, the particularity of offering it over wireless networks is considered. Indeed, the wireless channel is subject to a large variability due to fading, mobility, etc.  On top of this, it is a shared channel where multiple users are served simultaneously and cell capacity is divided among them. This introduces two variability time scales: flow level (tens of seconds) driven by the departures/arrivals of calls and wireless channel variability time scale (milliseconds) driven by the fast fading. In addition, the variable bit-rate (VBR) streaming leads to a variable service rate at
the time scale of tens of milliseconds.

\subsection{Related Literature}

Starting from the mid-nineties, many works focused on performance analysis for real time video delivery over wireless networks. A large attention was given to enhance video coding in order to combat errors introduced by the wireless channel variability. \cite{Stuhlmuller} derived a theoretical framework for the picture quality after video transmission over lossy channels, based on a 2-state Markov model describing burst errors on the symbol level. Authors in \cite{Zhang} and \cite{JSAC02:He} proposed methods for estimating the channel distortion and its impact on performance. These works mainly focused on ensuring robustness of video delivery over a variable wireless channel but did not consider the impact of flow level dynamics. A more recent set of works considered flow level performance in  cellular networks delivering real time video. Authors in \cite{Hegde} proposed a queuing theory model for deriving QoS when integrating elastic and video traffic in cellular networks; video QoS was expressed by a blocking rate, while average throughputs and delays represent QoS for elastic traffic. Authors in \cite{SalahTVT} derived the Erlang-like capacity region for a traffic mix including real time video, the aim being to dimension the network for ensuring a target QoS.  \cite{karrayTWireless} derived the stability region of the network and showed how it is impacted by real-time video traffic.

With the increased popularity of streaming services over wireless systems, more attention has been dedicated to deriving QoE performance metrics for this new streaming service, knowing the initial buffering period and its relationship with starvation. QoE issue has been addressed in the important works \cite{TMM10:Luan,TMM08:Liang,JSAC11:ParandehGheibi,Infocom12:Xu}. These works adopt different methodologies and assumptions for deriving QoE metrics. \cite{TMM10:Luan} considered a general G/G/1 queue where the arrival and service rates are characterized by their first two moments, while \cite{TMM08:Liang} considered a particular wireless channel model where the channel oscillates between \emph{good} and \emph{bad} states following the extended Gilbert model \cite{Sanneck}. Authors in \cite{JSAC11:ParandehGheibi} considered a particular P2P video streaming based on random linear network coding; this simplifies the packet requests at the network layer and allows to model the receiver buffer as an M/D/1 queue. Finally, an M/M/1 queue model has been adopted in \cite{Infocom12:Xu}, allowing to derive explicit formula for QoE metrics.

As of the tools used in the literature for deriving QoE metrics, they differ in the adopted system models. \cite{TMM10:Luan} adopted a diffusion approximation where the discrete buffer size is replaced with a Brownian motion whose drift and diffusion coefficients are calculated based on the first two moments of the arrival and service rates. \cite{TMM08:Liang} presented a probabilistic analysis based on an a priori knowledge of the playback and arrival curves. \cite{JSAC11:ParandehGheibi} calculated bounds on the playback interruption probability based on the adopted M/D/1 buffer model. Explicit formula of the exact distribution of the number of starvations has been obtained in \cite{Infocom12:Xu} based on a Ballot theorem approach \cite{Takacs}. Authors in \cite{Infocom12:Xu} also proposed an alternative approach for computing QoE metrics based on a recursive algorithm that performs better than the Ballot Theorem in terms of complexity. They further studied the
QoE metrics of a persistent video streaming in cellular networks in \cite{Networking12:Xu}.

The above-described works on QoE estimation are very useful for catching the impact of variability of the wireless channel due to fast fading or even user's mobility. However, the underlying models fail to capture the large variations due to flow dynamics. For instance, the diffusion approximation in \cite{TMM10:Luan} supposes that the drift and diffusion coefficients are constant over time, which is not true when the number of concurrent flows changes during playback in wireless environments. The assumption of Poisson packet arrivals in \cite{JSAC11:ParandehGheibi,Infocom12:Xu} also fails to take into account these flow dynamics. Note that the analysis of \cite{JSAC11:ParandehGheibi} has been generalized to a two-state Markovian arrival process, but this corresponds more to a bursty traffic due to a Gilbert channel model than to flow dynamics.

\subsection{Main Contributions and Organization}

To the best of our knowledge, this paper is the first attempt to assess the impact of flow dynamics on the
QoE of streaming. We model the system as two queues in tandem. The first queue, representing the scheduler
of the base station, is modeled as a processor sharing queue, while the second represents the playout buffer
whose arrival rates are governed by the output process of the base station queue. We first consider a static
channel (no fast fading) with Constant Bit Rate (CBR) streaming, and derive the prefetching delay distribution
and the starvation probability generation function using Partial Differential Equations (PDEs) as well as
Ordinary Differential Equations (ODEs) constructed over the Markov process describing the flow dynamics.
We then extend the model to the Variable Bit Rate (VBR) streaming using diffusion approximation. We next
extend the model to include a fast fading channel and show that the impact of flow dynamics is preponderant
over the variability of the channel due to fast fading. Extensive simulations show that our models are
accurate enough to be used in QoE prediction. Our analysis also sheds light on the novel QoE enhancement strategies.
The results presented here can be used by the base station to ``recommend'' the prefetching
parameters to the media player, and to guide the admission control and the scheduling algorithms.
The main contributions of this work are summarized as follows:
\begin{enumerate}
\item Developing an analytical framework for assessing the impact of flow dynamics  in wireless data networks on streaming QoE.
\item Evaluating the performance of both CBR and VBR streaming.
\item Showing that the variability of the throughput due to flow dynamics is preponderant over the impact of fast channel variability due to fast fading.
\end{enumerate}

The remainder of this paper is organized as follows. Section \ref{sec:motivation} describes the
system model and the QoE metrics. Section \ref{sec:cbr} presents the analytical framework
for analyzing QoE taking into account flow dynamics. VBR streaming is analyzed in Section
\ref{sec:vbr}. The analytical model is verified through simulations in section \ref{sec:simu}
and a perfect match is demonstrated. Section \ref{sec:extension} extends the QoE analysis
framework to include the impact of fast fading. It also shows how to analyze QoE in a general
case where streaming services coexist with classical data services. Section \ref{sec:conclusion} eventually concludes the paper.

\section{Problem Description and Model}
\label{sec:motivation}

In this section, we first describe our motivation and the network settings.
We then define the metrics of quality of experience
for media streaming service, and present a queueing model for the playout buffer at a user.

\subsection{Motivation and Network Description}

We consider a wireless data network that supports a number of flows.
When a new flow ``joins'' the network, it requests the streaming service from a media server.
After the connection has been built, the streaming packets are transmitted through the base station (BS).
The streaming flows have \emph{finite} sizes,
which means that a flow ``leaves'' the network once the transmission completes.
Note that each active user cannot watch more than one streams at the mobile device
simultaneously. Hence, we use the terms ``flow'' and ``user'' interchangeably.

In wireless data networks, a streaming flow may traverse both wired and wireless links,
whereas the BS is the
bottleneck for the sake of limited channel capacity.
In other words, the queue of an \emph{active} flow is always backlogged at the BS.
This assumption holds because most of Internet streaming servers
use TCP/HTTP protocols to deliver streaming packets. The
TCP protocol in the transport layer exploits the available
bandwidth by pumping as more packets as possible to the BS.
The BS can easily perform per-flow congestion control to limit TCP
sending rate to avoid buffer overflow (a small number of concurrent flows in total).
The adaptive coding and modulation in the physical layer, and ARQ scheme at the MAC layer
can effectively avoid TCP packet loss. Due to these reasons, we do not consider TCP packet losses in our system.

%
%
Streaming flows may experience fast fading and normalized signal-to-noise ratio (NSNR) scheduling is usually adopted to achieve multiuser diversity with
the consideration of fairness \cite{TVT07:Choi,TOC06:Song}.
The scheduling duration is commonly around 2ms \cite{Bonald1}.
NSNR selects the user that has the largest ratio of SNR compared with its mean SNR.
It is similar to the well-known proportional fair (PF) scheduler
in that they both attempt to achieve channel access-time fairness.
We consider NSNR instead of
PF for two reasons. First, the moments of throughput of PF
do not have explicit results, even asymptotic ones (see \cite{TOC06:Song} and references therein)
when the channel capacity is computed according to the Shannon theorem.
Second, NSNR needs the knowledge of the average SNR that can be obtained from the history information.
When a flow join the network, its throughput process is stationary as long as the number
of active flows does not change. However, the throughput of PF scheduler is not
stationary, but is a dynamic function of time $t$ (see \cite{TWC04:Whiting} for the ODE throughput model
with two users). It relies on the configuration of the average throughput at time 0.
The initial average throughput may influence the start-up delay, and cause the whole system
intractable. Here, we make a declaration that our analytical framework applies
to any wireless scheduling algorithm whose first two moments of throughput 
per-slot can be derived.




At the user side, incoming bits are reassembled into video \emph{frames}
step by step. These video frames are played with a deterministic rate, e.g. 25 frames per second (fps)
in the TV and movie-making businesses.
The size of a frame is determined by the video codec, i.e. a high definition video streaming
or a complex video scenario require more bits to render each frame.
We consider two modes of streaming services: constant bit-rate (CBR) and variable bit-rate (VBR).
In CBR, the rate at which a codec's output data should be consumed is constant (i.e. the
same size of frames). The VBR streaming has a variable frame size so as to deliver a more efficiently encoded
and consistent watching experience. The frame size roughly follows Erlang/Gamma distributions \cite{MASI08}.

\begin{figure}[!htb]
    \centering
    \includegraphics[width=3.0in, height=0.7in]{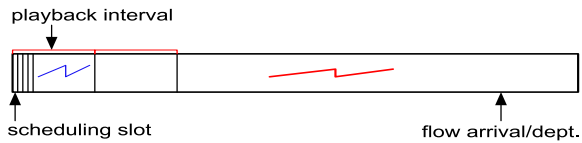}
    \caption{Illustration of three different time scales}
    \label{fig:time_scales}
\end{figure}

We highlight the properties of the streaming system briefly to facilitate the mathematical modeling.
In our system, there exist three time scales shown in Fig.\ref{fig:time_scales}: i) the scheduling duration (e.g. 2ms); ii)
playback interval (e.g. 40ms for a video frame rate of 25fps), and iii) duration of flow dynamics (lasting about tens of seconds).
The scheduler and the media player do not work at the same granularity of time scale and job size.

\subsection{QoE Metrics}

There exist five industry-standard video quality metrics. Authors in \cite{Sigcomm2011} summarize them
into five terms: \emph{join time}, \emph{buffering ratio}, \emph{rate of buffering events},
\emph{average bitrate} and \emph{rendering quality}. The first three metrics reflect
the fundamental tradeoff in designing the prefetching process. The last two metrics
are concerned with source coding. For
analytical convenience, we redefine the QoE metrics regarding ``prefetching'' process.

\noindent
- \textbf{Start-up delay:} The start-up delay denotes the duration (measured in seconds)
between the time that a user initiates a session and the time that the media player starts
playing video frames. In the initial prefetching phase, the player starts until
the duration of received video reaches the \emph{start-up threshold} measured in seconds of video segment.
The start-up delay depicts the user's impatience of waiting for the
video playback. Once the starvation event happens, the player pauses and resumes until
the rebuffered video duration reaches the \emph{rebuffering threshold}. We use the term
\emph{rebuffering delay} to differentiate the rebuffering time from the initial start-up delay.

\noindent
- \textbf{Starvation probabilities:} When
the playout buffer of a user becomes empty
before the video has been completely played, we call this event a \emph{starvation}.
The starvation is very annoying to users. We adopt the starvation probability to evaluate
the influence of the start-up threshold. In addition, if the rebuffering process is taken
into account, we analyze the probabilities of having a certain number of starvations.

Note that the start-up delay and the starvation probabilities can be used to compute the
QoE metrics in \cite{Sigcomm2011}. The expected number of starvations
is the sum of the products of the number of starvations and its probability. The expected
buffering time equals to the product of the start-up delay in each rebuffering and
the mean number of starvation events (including
the initial prefetching).

\subsection{Basic Queueing Model of Playout Buffer}

We consider a wireless cellular network that supports up to $K$ simultaneous flows.
The purpose of admission control is to avoid the overloading of the cell.
We make the following assumptions:

\noindent
- \textbf{Single user type and static channel:} We begin with the case where streaming users coexist in a static channel, as this provides an easier route to understand
the developed QoE evaluation model. The impact of fast fading is added in section \ref{sec:extension}. We also consider that all the flows have the same SNR, and hence, in a static channel case, identical throughput.
The extension to multiple user classes is presented in Section \ref{sec:conclusion}.


\noindent
- \textbf{Exponentially distributed video duration:} The video duration, measured in
seconds, is exponentially distributed with mean $1/\theta$.
Though the exponential distribution is not the most realistic way to describe video duration,
it reveals the essential features of the system, and is the first
step for more general distributions.
Later on (in section \ref{sec:conclusion}), we allow the video length to have the
hyper-exponentially distribution that is commonly adopted in wireless networks \cite{TON08:Salah}.

\noindent
- \textbf{Processor sharing at the BS:} The scheduling slot is
very small (e.g. $\leq$2ms in 3G LTE) compared with the service interval between
two video frames (e.g. 40ms at 25fps) in the playout buffer.
This property enables us to treat the BS as an egalitarian processor sharing queue where
all the flows are served simultaneously. Hence, the per-flow throughput, depicted in continuous time,
is a deterministic step-wise function of the number of active users in the static channel (e.g. \cite{Borst}).
%
%

\noindent
- \textbf{Continuous time playback:} The service of video contents is regarded as a continuous
process, instead of a discrete rendering of adjacent video frames spaced by a fixed interval.
This assumption is commonly used (see \cite{wangbing}) and is validated by simulations in this work.


%

We denote by $\lambda$ the arrival rate of new video streams.
Let $Bitrate$ be the playback speed of video streams in bits per-second, and $C$ (in bps) be the capacity of the static wireless channel.
Given the exponential distribution of video duration, the file size $F$ (measured in bits) is also exponentially
distribution with mean $1/\theta_F=Bitrate/\theta$. Therefore, the dynamics of coexisting flows in the cell
can be depicted as a continuous time Markov chain with a finite state space.

We concentrate on one ``tagged'' flow in order to gain the insight of dynamics of the playout buffer.
At any time $t$,
the tagged flow sees $i$ other flows in a finite space $S:=\{0,1,\cdots,K-1\}$. We denote by $\{I(t);t\geq 0\}$
the external environment process that influences the throughput of the tagged flow.
The environmental change refers to the join of a new flow, or the departure of an existing flow.
From the assumption
of Poisson flow arrival and exponentially distributed flow size, we can see that $\{I(t);t\geq 0\}$
is a homogeneous, irreducible and recurrent Markov process.
Let $\{\pi_i;i\in S\}$ be the stationary
distribution of environmental states that will be computed in the following sections.
The throughput of the tagger user is $b_i:=\frac{C}{Bitrate\cdot(i+1)}$
in seconds of video contents at state $i$. Let $N_e(t)$ be the number of changes in
the environment by time $t$. Denote by $A_l$ the time that the $l^{th}$ environmental change takes place with $A_0 = 0$
and by $I_{l}:=I(A_l)$ the state to which the environment changes after time $A_l$.
When the tagged flow joins the network, we begin to study the dynamics of its playout buffer length.
The entry time of the tagged flow is set to $t=0$.

We denote by $Q(t)$ the length of playout buffer \emph{measured in seconds} of video contents at time $t$. In the prefetching phase,
$Q(t)$ is expressed as
\begin{eqnarray}
Q_a(t) = \sum_{l=1}^{N_e(t)} b_{I_l}(A_l - A_{l-1}) + b_{I_{N_e(t)}}(t-A_{N_e(t)}).
\label{eq:queue_startup}
\end{eqnarray}
Denote by $q_a$ the start-up threshold. The start-up delay $T_a$ is defined as
\begin{eqnarray}
T_a = \inf\{t\geq 0| Q_a(t) \geq q_a\}.
\label{eq:define_startupdelay}
\end{eqnarray}
The cumulative distribution of $T_a$ is expressed as
\begin{eqnarray}
\Psi_i(t;q_a) = \mathbb{P}\{T_a < t|I(0) = i\}
\label{eq:distribution_startupdelay}
\end{eqnarray}
if the tagged flow is in state $i$ upon arrival.

Let $q$ be the duration of buffered video content in seconds before the video playback.
When the media player starts the rendering, the queueing process $\{Q(t); t\geq 0\}$ is given by
\begin{eqnarray}
Q_b(t) = q - t {+} \sum_{l=1}^{N_e(t)} b_{I_l}(A_l {-} A_{l-1}) + b_{I_{N_e(t)}}(t{-}A_{N_e(t)}),
\label{eq:queue_service}
\end{eqnarray}
if the time axis starts at the instant of playing. Define $c_i := b_i - 1$ for all $i\in S$. Define
\begin{eqnarray}
T_b = \inf\{t\geq 0| Q_b(t) < 0\}
\label{eq:define_ruintime}
\end{eqnarray}
to be the time of observing empty buffer. Denote by $T_e (T_e<\infty)$ the completion time of downloading of
the tagged flow. If $T_b$ is less than $T_e$, a starvation event happens at the playout buffer.
Then, the ultimate starvation probability is computed as
\begin{eqnarray}
W_i(q_a) = \mathbb{P}\{T_b < T_e|I(0) = i, Q_b(0) = q_a\}
\label{eq:distribution_ruintime}
\end{eqnarray}
when the playback begins at state $i$, and stops at an arbitrary 
state that meets an empty queue for the first time.
The ultimate starvation probability is the weighted sum of starvation probabilities
at all the ergodic entry states.

\section{Complete QoE analysis for CBR streaming}
\label{sec:cbr}

In this section, we model the starvation probability and the prefetching
delay in a static channel
where the media flows join and leave the system dynamically. The key idea
is to investigate the queueing process of one ``tagged'' flow on the basis
of differential equations.

\subsection{Markov models of flow dynamics}

Our purpose here is to construct two Markov chains to characterize the dynamics
of the number of active flows. The first one models flow dynamics
before the ``tagged'' flow joins in the network. Based on this Markov process,
we can compute the stationary distribution of the number of active flows
observed by the ``tagged'' flow at the instant when it is admitted.
The second one describes the flow dynamics after the tagged flow is admitted.
This Markov process enables us to investigate how the playout buffer of the tagged user
changes.

\begin{figure}[!htb]
    \centering
    \includegraphics[width=3.2in]{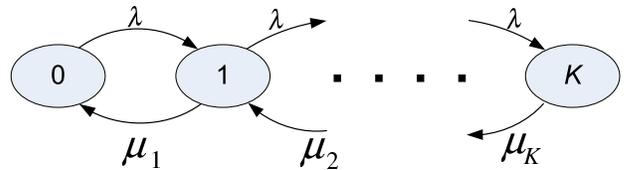}
    \caption{Markov chain before the tagged flow joins}
    \label{fig:markovchain0}
\end{figure}

We first look into the flow dynamics before the tagged flow joins.
When the NSNR scheduling
algorithm is used, the per-flow throughput is proportional to the reciprocal
of flow population. Given the Poisson arrival rate and the exponentially
distributed service time, we can model the flow dynamics
as a finite-state Markov chain $\mathbf{Z}_a:=\{0,1,\cdots,K\}$ shown in Fig.\ref{fig:markovchain0}.
The transition rate from $i$ to $i-1$ is $\mu_i:=C\theta_F$.
Note that the network capacity is a constant in the static channel.
Hence, we let $\mu_i = \mu$ for $i=1,2,\cdots, K$ and $\mu_0 = 0$.
Define $\rho:=\frac{\lambda}{\mu}$ to be the load of the channel.
Let $z_i^a$ be the stationary probability that there exist $i$ flows. We give
the expression of $z_i^a \;(i\in S\cup\{K\})$ directly because it is easy to compute.
\begin{eqnarray}
z_0^a = \frac{1-\rho}{1-\rho^{K{+}1}};\;\;\;\; z_i^a = \frac{\rho^i(1-\rho)}{1-\rho^{K{+}1}},\;\;\; \forall i=1,\cdots, K. \nonumber
\label{eq:stationarydistribution0}
\end{eqnarray}
The tagged user cannot be admitted at state $K$ due to the admission control at the BS. Therefore, if it joins
in the network successfully, it will observe $i$ other flows with the probability $\pi_i$,
\begin{eqnarray}
\pi_i = \frac{z_i^a}{1-z_K^a} = \frac{\rho^i(1-\rho)}{1-\rho^K}, \;\;\; \forall i\in S.
\label{eq:stationarydistribution1}
\end{eqnarray}

\begin{figure}[!htb]
    \centering
    \includegraphics[width=3.2in]{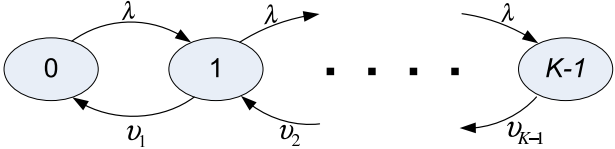}
    \caption{Flow dynamics observed by tagged flow}
    \label{fig:markovchain1}
\end{figure}

After the tagged flow joins in the network, the Markov process $\mathbf{Z}_a$ has been altered.
The states are the number of flows observed by the tagged user, and the transition rates
are conditioned on the presence of the tagged flow.
Therefore, we model the flow dynamics observed by the tagged flow
through a finite-state Markov chain $\mathbf{Z}_b:=\{0,1,\cdots,K{-}1\}$ in Fig.\ref{fig:markovchain1}.
Denoted by $\nu_i$ the transition rate from state $i$ to $i{-}1$. The per-flow throughput at state $i$
is $\frac{C}{(i{+}1)}$ so that there has $\nu_i:=\frac{iC\theta_F}{(i{+}1)} = \frac{i}{i{+}1}\mu$
for all $i\in S$.
For the simplicity of notations, we denote by $\lambda_i$ the transition rate from state $i$ to $i+1$.
It is obvious to have $\lambda_i=\lambda$ for all $i\neq K{-}1$ and $\lambda_{K{-}1} = 0$.

\subsection{Modeling prefetching delay distribution}
\label{subsec:prefetching}

We want to know how long the tagged user needs to wait in the prefetching phase.
Recall that $q_a$ is the start-up threshold.
The prefetching time is only meaningful to the case that the video duration
is \textbf{longer} than $q_a$.
In the prefetching phase, because the playout buffer
does not serve video frames, the queue length of the tagged flow evolves in
an infinitesimal time interval $[0,h]$ with $h (>0)$
\begin{eqnarray}
Q(t+h) = Q(t) + b_ih.
\label{eq:prefetchingqueue1}
\end{eqnarray}
The distribution of the prefetching time is difficult to solve directly.
We resort to the following duality problem:

\medskip
\boxed{
\begin{minipage}{3.1in}
{\ensuremath{\mbox{\sc  Duality Problem:}}}
What is the starvation probability by time $t$ if the queue is depleted with rate
$b_i (i\in S)$ and the duration of prefetched contents is $q_a$?
\end{minipage}
}
\medskip

In the duality problem, the queue dynamics in $[0, h]$ is modified as
\begin{eqnarray}
\tilde{Q}(t+h) = \tilde{Q}(t) - b_ih.
\label{eq:prefetchingqueue2}
\end{eqnarray}

We define $U_i(q,t)$ $(\forall i\in S)$ to be the probability of starvation
before time $t$, conditioned on the entry state $i$ and the initially prefetched
content $q$. We use differential equations to obtain $U_i(q,t)$.
In the infinitesimal time interval $[0,h]$, there are four possible events
\begin{itemize}

\item no change of the concurrent flows;

\item arrival of one flow;

\item departure of one flow (not the tagged one);

\item occurrence of more than one events.
\end{itemize}
\noindent Conditioned on the events occurred in $[0,h]$, we have
\begin{eqnarray}
U_i(q,t) \!\!\!&=&\!\!\! (1-\lambda_i h-\nu_ih) U_i(q-b_ih, t-h) \nonumber\\
\!\!\!&&\!\!\!\!\!\!\!\!\!\!\!\!\!\!\!\!\!\! + \lambda_i h U_{i+1}(q-b_ih, t-h) \nonumber\\
\!\!\!&&\!\!\!\!\!\!\!\!\!\!\!\!\!\!\!\!\!\! + \nu_i h U_{i-1}(q-b_ih, t-h) + o(h), \;\; \forall i\in S.
\label{eq:solving_startupdelay_eq1}
\end{eqnarray}
The above equation yields for $i\in S$
\begin{eqnarray}
&&\!\!\!\!\!\!\!\!\!\!\!\!\frac{1}{h}(U_i(q,t) - U_i(q-b_ih, t-h)) = -(\lambda_i+\nu_i)U_i(q-b_ih, t-h) \nonumber\\
&&\!\!\!\!\!\!\!\!\!\!\!\! + \lambda_i  U_{i+1}(q{-}b_ih, t{-}h) + \nu_i U_{i-1}(q{-}b_ih, t{-}h) +  o(h)/h.
\label{eq:solving_startupdelay_eq2}
\end{eqnarray}
When $h\rightarrow 0$, the left side of eq.\eqref{eq:solving_startupdelay_eq2}
is the partial differentials of $U_i(q,t)$ over $q$ and $t$. In other words, eq.\eqref{eq:solving_startupdelay_eq2}
yields a set of linear partial differential equations (PDEs)
\begin{eqnarray}
\frac{\partial U_i}{\partial t}\!\!&=&\!\! -b_i \frac{\partial U_i}{\partial q} -(\lambda_i+\nu_i)U_i(q,t) \nonumber\\
&& + \lambda_i  U_{i+1}(q,t) + \nu_i U_{i-1}(q,t), \;\; \forall i\in S,
\label{eq:solving_startupdelay_eq3}
\end{eqnarray}
with the initial condition
\begin{eqnarray}
U_i(q,0) = 0, \quad \forall q > 0
  \label{eq:solving_startupdelay_eq4}
\end{eqnarray}
and the boundary conditions at both sides
\begin{eqnarray}
U_i(0,t) \!\!&=&\!\! 1,\;\;\; \forall \; t\geq 0,
\label{eq:solving_startupdelay_eq5}\\
\lim_{q\rightarrow \infty} U_i(q,t) \!\!&=&\!\! 0,\;\;\; \forall \; t\geq 0.
\label{eq:solving_startupdelay_eq6}
\end{eqnarray}
The initial condition in eq.\eqref{eq:solving_startupdelay_eq4} means
that the starvation cannot happen at time 0 for $q > 0$.
The right-side boundary condition says that the starvation will not happen
before $t$ if the initial prefetching is large enough.
Comparing Eq.\eqref{eq:solving_startupdelay_eq4} with Eq.\eqref{eq:solving_startupdelay_eq5},
we find that $U_i(q,t)$ is discontinuous at $(q,t)=(0,0)$. 
This greatly increases the complexity to obtain $U_i(q,t)$, which will be shown later. 
Here, the c.d.f. of start-up delay is the solution of linear PDEs
by letting $q$ be $q_a$.  To solve the linear PDEs, we first define a matrix as
\begin{eqnarray}
\mathbf{M}_{S}=\left( \begin{array}{cccccc}
\lambda_0 & -\lambda_0 & 0 & \cdots & 0 & 0 \\
-\nu_1 & \lambda_1+\nu_1 & -\lambda_1 & \cdots & 0 & 0 \\
\cdots & \cdots & \cdots & \cdots & \cdots & \cdots \\
0 & 0 & \cdots & \cdots & -\nu_{N{-}1} & \nu_{N{-}1} \end{array} \right).
\label{eq:eq_mus}
\end{eqnarray}
According to the lemma in Appendix.B, the tridiagonal matrix $\mathbf{M}_{S}$ is diagonizable
Let $D_S$ be an invertible matrix, and $\Lambda_S$ be a diagonal matrix that contains the eigenvalues 
of $\mathbf{M}_{S}$. Then, there has $\mathbf{M}_{S}=D_S\Lambda_SD_S^{-1}$. 
Define a vector function $\mathbf{F}(q,t)$ as
\begin{eqnarray}
\mathbf{F}_i(q,t) {=} 1-\Phi(\frac{q-b_it}{\sqrt{\alpha t}}) 
, \quad \forall i\in \mathbf{S},
\label{eq:g_function}
\end{eqnarray}
where $\alpha$ is a very small positive and $\Phi(x) = (1/\sqrt{2\pi})\int_{-\infty}^{x}e^{-y^2/2}dy=\frac{1}{2}\textbf{erfc}(-\frac{x}{\sqrt{2}})$.
Then, the linear PDEs in Eq.\eqref{eq:solving_startupdelay_eq3} are solved by
\begin{eqnarray}
\mathbf{U}(q,t) = D_S\exp{(-\Lambda_St)}D_S^{-1}\cdot\{1-\frac{1}{2}\textbf{erfc}(-\frac{q{-}b_it}{\sqrt{2\alpha t}})\}.
\label{eq:pde_solution}
\end{eqnarray}
So far, we have derived the explicit c.d.f. of start-up delay, which only involves a small-scale matrix decomposition.
Detailed analysis can be found in Appendix. 

\noindent \textbf{Remark:} The numerical integral of the PDEs may be unstable due to the discontinuity at the point 
$(q,t)=(0,0)$. The approximated model using Brownian motion offers a close-form expression, while is less accurate
than the numerical integral.

We next analyze the probability that the prefetching process starts at state $i$ and ends at state $j$, for all $i,j\in S$.
Define
\begin{eqnarray}
V_{i,j}(q;q_a) := \mathbb{P}\{I(T_a)=j|I(0)=i,Q(0)=q\}.
\label{eq:solving_startupprocess1}
\end{eqnarray}
\noindent We can use the approach of obtaining $U_i(q,t)$ to solve $V_{i,j}(q;q_a)$.
Note that we now use the queueing dynamics in eq.\eqref{eq:prefetchingqueue1} instead of eq.\eqref{eq:prefetchingqueue2}.
In the time interval $[0,h]$, there exists for all $i,j\in S$
\begin{eqnarray}
&&\!\!\!\!\!\!\!\!\!\!\!\!\!V_{i,j}(q;q_a) = (1-\lambda_ih-\nu_ih)V_{i,j}(q+b_ih;q_a)  \nonumber\\
&& \!\!\!\!\!\!\!\!+ \lambda_ihV_{i{+}1,j}(q{+}b_ih;q_a)+ \nu_i hV_{i{-}1,j}(q{+}b_ih;q_a)+ o(h).
\label{eq:solving_startupprocess2}
\end{eqnarray}
It is easy to see that $V_{i,j}(q;q_a)$ is the solution of the following differential equation
\begin{eqnarray}
b_i\dot{V}_{i,j}(q;q_a) \!\!\!&=&\!\!\! (\lambda_i+\nu_i)V_{i,j}(q;q_a) - \lambda_iV_{i{+}1,j}(q;q_a) \nonumber\\
&& - \nu_i V_{i{-}1,j}(q;q_a), \;\forall i,j\in S,
\label{eq:solving_startupprocess2}
\end{eqnarray}
with the boundary condition
\begin{eqnarray}
V_{i,j}(q_a;q_a) := \left\{\begin{matrix}
\;1 \;\;\; &&\textrm{ if }\; i=j ;\\
\;0 \;\;\; &&\;\;\textrm{otherwise }.
\end{matrix}\right.
\label{eq:solving_startupprocess3}
\end{eqnarray}
We interpret the boundary condition in the following way. If there exist $I(0)=i$ and $Q(0)=q_a$,
the prefetching duration is 0 and the prefetching process ends at state $i$.
Hence, $V_{i,j}(q_a;q_a)$ is 1 iff $i$ equals to $j$. 
Define a matrix $\mathbf{M}_{V}$ as $\mathbf{M}_{V} = \mathbf{diag}\{\frac{1}{b_i}\}\cdot\mathbf{M}_S$. 
We have the following property w.r.t the eigenvalues of $\mathbf{M}_{V}$.
\begin{lemma}
\label{lemma:no1} The matrix $\mathbf{M}_{V}$ has $K$ real non-negative eigenvalues, and is similar to a diagonal matrix.
\end{lemma}
\noindent Define $\mathbf{1}_j$ to be a column vector in which the $j^{th}$ element is 1 and all other elements are 0.
Eq. \eqref{eq:solving_startupprocess2} can be rewritten as
\begin{eqnarray}
\mathbf{\dot{V}}(q;q_a) = \mathbf{M}_{V} \mathbf{V}(q;q_a).
\label{eq:solving_startupprocess3.5}
\end{eqnarray}
Then, $\mathbf{V}(q;q_a)$ is solved by 
\begin{eqnarray}
\mathbf{V}(q;q_a) = \exp{(\mathbf{M}_{V}q)}\cdot \mathbf{V}(0;q_a).
\label{eq:solving_startupprocess3.6}
\end{eqnarray}
According to Lemma \ref{lemma:no1}, we let $\mathbf{M}_{V} :=D_{V}\Lambda_{V}D_{V}^{-1}$ where $D_{V}$ is an invertible matrix and
$\Lambda_{V}$ is the diagonal matrix containing all the eigenvalues of $\mathbf{M}_{V} $. Therefore, Eq.\eqref{eq:solving_startupprocess3.6}
is expressed as 
\begin{eqnarray}
\mathbf{V}(q;q_a) = D_V\exp{(\Lambda_{V}q)}D_V^{-1}\cdot \mathbf{V}(0;q_a).
\label{eq:solving_startupprocess3.6}
\end{eqnarray}
Submitting eq.\eqref{eq:solving_startupprocess3} to eq.\eqref{eq:solving_startupprocess3.6}, we yield
\begin{eqnarray}
\mathbf{V}(q;q_a) = D_V\exp{(\Lambda_{V}(q-q_a))}D_V^{-1}\cdot \mathbf{V}(q_a;q_a).
\label{eq:solving_startupprocess3.7}
\end{eqnarray}

\subsection{Modeling starvation probability}

\begin{figure}[!htb]
    \centering
    \includegraphics[width=3.2in]{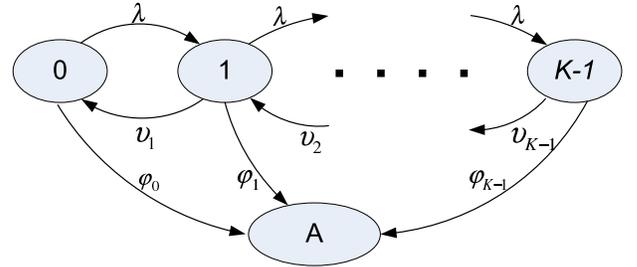}
    \caption{Markov chain for user dynamics with an
    absorbing state for departure of tagged flow}
    \label{fig:markovchain2}
\end{figure}

The modeling of starvation probabilities should
take into account the departure of the tagged flow. Recall that
the CTMC in Fig. \ref{fig:markovchain1} assumes the persistent
tagged flow, which is not suitable for the playback process.
Before solving the starvation probabilities,
we first modify the original CTMC by adding an absorbing state \textbf{A} shown in Fig. \ref{fig:markovchain2}.
The state \textbf{A} denotes the event that the tagged flow completes its downloading.
Because of the exponentially distributed video duration,
the transition from state $i$ to
state \textbf{A} is Poisson. Denote by $\varphi_i$ the transition
rate from state $i$ to \textbf{A}. At state $i$, the bandwidth of a flow
is $\frac{C}{i+1}$, resulting in $\varphi_i := \frac{\mu}{i+1}$.
Define $c_i := b_i - 1$.
The queue length of the tagged flow changes in an infinitesimal interval $h$ according to the rule
\begin{eqnarray}
Q(t+h) = Q(t) + c_ih.
\label{eq:queuedyn_playback}
\end{eqnarray}
If $c_i>0$, the bandwidth is sufficient for continuous playback of the tagged flow and $i$ other flows.
For mathematical convenience, we suppose
that $q$ is $0^{-}$ if buffer starvation happens.
 When the tagged flow enters the absorbing state, it has downloaded the whole file with a non-empty playout
buffer. Thus, the starvation probability at state \textbf{A} is 0 for any $q\geq 0$.
Let $W_i(q)$ be the starvation probability with $q$ seconds of contents in the playout buffer at state $i$.
We derive a system of ordinary differential equations for $W_i(q)$. In an
infinitesimal interval $[0,h]$, there are five possible events:
\begin{itemize}

\item no change of the concurrent flows;

\item arrival of one more flow;

\item departure of one flow (not the tagged flow);

\item the tagged flow entering the absorbing state;

\item occurrence of more than one events.
\end{itemize}
The above conditions give rise to the a set of equations
\begin{eqnarray}
\!\!\!\!W_i(q) \!\!\!&=&\!\!\! (1-(\lambda_i+\mu_i)h)W_i(q+c_ih) \nonumber\\
\!\!\!\!\!\!\!\!\!\!\!&&\!\!\! +\lambda_i W_{i+1}(q+c_ih)  + \nu_{i} W_{i-1}(q+c_ih) +o(h).
\label{eq:solving_starveprob1}
\end{eqnarray}
\noindent When $h\rightarrow 0$, we obtain
\begin{eqnarray}
c_i\dot{W}_i(q) = (\lambda_i+\mu_i)W_i(q)  -\lambda_i W_{i+1}(q)  - \nu_{i} W_{i-1}(q).
\label{eq:solving_starveprob1}
\end{eqnarray}
The above equations can be rewritten in the matrix form
\begin{eqnarray}
\mathbf{\dot{W}}(q) = \mathbf{M}_W\mathbf{W}(q)
\label{eq:matrixform1}
\end{eqnarray}
\noindent where $\mathbf{M}_W$ is expressed in eq.\eqref{eq:matrixform2}
\begin{eqnarray}
\left( \begin{array}{cccccc}
\frac{\lambda_0+\mu_0}{c_0} & -\frac{\lambda_0}{c_0} & 0 & \cdots & 0 & 0 \\
-\frac{\nu_1}{c_1} & \frac{\lambda_1+\mu_1}{c_1} & -\frac{\lambda_1}{c_1} & \cdots & 0 & 0 \\
\cdots & \cdots & \cdots & \cdots & \cdots & \cdots \\
0 & 0 & \cdots & \cdots & -\frac{\nu_{N{-}1}}{c_{N{-}1}} & \frac{\mu_{N{-}1}+\lambda_{N-1}}{c_{N{-}1}} \end{array} \right).
\label{eq:matrixform2}
\end{eqnarray}
The solution to eq.\eqref{eq:matrixform1} is given directly by
\begin{eqnarray}
\mathbf{W}(q) =\exp{ (\mathbf{M}_Wq)}\cdot\mathbf{W}(0),
\label{eq:matrixform2.1}
\end{eqnarray}
where $\mathbf{W}(0)$ denotes the starvation probabilities with no initial prefetching. 
The boundary conditions are $W_i(q)=0$ for all $i$ as $q$ approaches infinity.
Note that $W_i(0)=1$ holds for all $i$ if $c_i<0$. Otherwise, $W_i(0)$ are unknowns for all $i$ with $c_i>0$. 
Using the proof of Lemma \ref{lemma:no1}, we can show that $\mathbf{M}_W$ is similar to a diagonal matrix.
There exist an invertible matrix $D_W$ and a diagonal matrix $\Lambda_W$ such that 
$\mathbf{M}_W:=D_W\Lambda_WD_W^{-1}$.
The starvation probabilities $\mathbf{W}(q) $ are expressed as
\begin{eqnarray}
\mathbf{W}(q) = D_W\exp{(\Lambda_Wq)}D_W^{-1}\cdot\mathbf{W}(0).
\label{eq:matrixform2.1}
\end{eqnarray}
The eigenvalues in $\Lambda_W$ are sorted in a decreasing order.
According to Gershgorin circle theorem \cite{matrixbook}, 
the signs of eigenvalues are uncertain since the centers of the Gershgorin
circles can be positive or negative. Based on the signs of $c_i$ for $i\in\mathbf{S}$, we obtain the following corollary. 
\begin{corollary}
Suppose that $c_i$ is positive for $0\leq i<k$ and is negative for $k\leq i <K$.
The matrix $\mathbf{M}_W$ has $k$ positive eigenvalues and $K{-}k$ negative eigenvalues.
\end{corollary}
The unknowns in $\mathbf{W}(0)$ can be solved subsequently. 
Define a vector $\bar{\mathbf{W}}:=D_W^{-1}\cdot\mathbf{W}(0)$. 
When $q$ is infinitely large, $\mathbf{W}(q)$ is a zero vector, resulting in 
$\exp{(\Lambda_Wq)}D_W^{-1}\cdot\mathbf{W}(0) = 0$.
Because the first $k$ eigenvalues are positive in $\Lambda_W$, there must have 
$\bar{W}_i=0$ for $i<k$. Hence, the unknowns $W_i(0)$ for $i<k$ can be derived.

Next, we build a bridge to interconnect the prefetching
threshold and the starvation probability function $W_i(q)$.
For a given prefetching threshold $q_a$,
the starvation event takes place only when the video duration $T_{video}$ is longer than $q_a$.
This is to say, a flow with $T_{video} > q_a$ can be regarded as a tagged flow.
When the prefetching process is finished, the tagged flow enters the playback process.
Conditioned on the distribution of entry states $\mathbf{\pi}$,
the distribution of the states that the playback process begins (or the prefetching process ends) is computed by
$\mathbf{\pi}\cdot \mathbf{V}(0;q_a)$.
Then, the starvation probability with the prefetching threshold $q_a$ is obtained by
\begin{eqnarray}
P_s(q_a) &=& \mathbb{P}\{T_{video} > q_a\}\cdot \mathbf{\pi}\cdot \mathbf{V}(0;q_a)\cdot\mathbf{W}(q_a) \nonumber\\
&=& \exp\big(-\theta q_a\big)\cdot \mathbf{\pi}\cdot \mathbf{V}(0;q_a)\cdot\mathbf{W}(q_a).
\end{eqnarray}

\subsection{Modeling P.G.F. of starvation events}

When a starvation event happens, the media player pauses until $q_b$ seconds of video contents are
re-buffered. A more interesting but challenging problem is how many starvations may happen in a streaming
session. In this section, we come up with an approach to derive the probability generating function
of starvation events.

\begin{figure*}[!htb]
    \centering
    \includegraphics[width=6in]{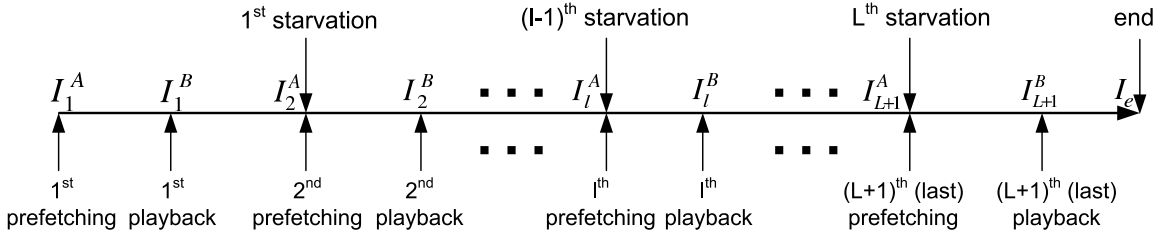}
    \caption{A path with $L$ starvations}
    \label{fig:samplepath}
\end{figure*}

We define a \emph{path} as a sequence of prefetching and starvation events, as well as the
event of completing the downloading. Obviously, the probability of
a path depends on the number of starvations. We illustrate a
typical path with $L$ starvations in figure \ref{fig:samplepath}
that starts from a prefetching process and ends at a playback process.
We denote by $I_{l}^A$ the beginning state of the
$l^{th}$ prefetching, by $I_{l}^B$ the beginning state of the $l^{th}$ playback,
and by $I_{e}$ the end of downloading.
The end of a prefetching process
is exactly the beginning of a playback process. The end of a playback process
is also the beginning of a subsequent prefetching process
if the video has not been downloaded completely.
This path contains a sequence of events happening at the states
$\{I_1^A,I_1^B,I_2^A,I_2^B,\cdots, I_{L+1}^A, I_{L+1}^B, I_{e}\}$. The process between $I_l^A$ and $I_l^B$
is the $l^{th}$ prefetching process, while that between $I_l^B$ and $I_{l+1}^A$ is the
$l^{th}$ playback process, ($1\leq l\leq L$). The first starvation
takes place at the instant that the second prefetching process begins.
The starvation event (e.g. $I_l^B,\; 1\leq l\leq L$) cannot happen at the state $i$ that
has $c_i\geq 0$.


The sample path in figure \ref{fig:samplepath} demonstrates a roadmap to find the p.g.f. of starvation events.
We need to compute the transition probability along the path with all possible states.
Recall that the transition probabilities from state $I_l^A$ to $I_l^B$ have been computed
in section \ref{subsec:prefetching}. The only missing
part is the transition probabilities from state $I_l^B$ to $I_{l+1}^A$.

Denote by $X_{i,j}(q)$ the probability that a playback process starts at
state $i$ and meets with the empty buffer at state $j$ with the prefetching threshold $q$.
Define a matrix $\mathbf{X}(q):=\{X_{i,j}(q);i,j\in S\}$.
Denote by $\mathbf{X}_{j}(q)$ the vector of probabilities that the starvation takes place at state $j$
with the prefetching threshold $q$, i.e. ${\small\mathbf{X}_{j}(q):=[X_{0,j}(q), \cdots, X_{K{-}1,j}(q)]^T}$.
Let ${\small\mathbf{X}_{j}(0):=[X_{0,j}(0),\cdots, X_{K{-}1,j}(0)]^T}$ be the vector of those probabilities without
the prefetching.
Using the same argument, we get the differential equation of $X_{i,j}(q)$, $\forall i,j\in S$,
\begin{eqnarray}
c_i\dot{X}_{i,j}(q) = (\lambda_i+\mu_i)X_{i,j}(q) {-}\lambda_i X_{i+1,j}(q)  {-} \nu_{i} X_{i-1,j}(q).
\label{eq:starvation_pgf1}
\end{eqnarray}
The solution of eq.\eqref{eq:starvation_pgf1} is directly given by
\begin{eqnarray}
\mathbf{X}_j(q) = D_W\exp{(\Lambda_Wq)}D_W^{-1}\cdot \mathbf{X}_j(0).
\label{eq:starvation_pgf2}
\end{eqnarray}
The computation of $\mathbf{X}_j(q)$ requires the knowledge of the boundary condition $\mathbf{X}_j(0)$.
Here, $X_{i,j}(0) = 0,\; i\neq j$ and $X_{i,j}(0) = 1$ if $c_i < 0$, and 
$X_{i,j}(0) = 0$ if $c_{K{-}1}\geq 0$. The computation of remaining $X_{i,j}(0)$ follows the same approach 
as that in the computation of $W_i(0)$.

When replacing $q$ by $q_a$, we obtain the probability $X_{ij}(q_a)$ that the first starvation
happens at state $j$ with $i$ other flows observed by the tagged flow at the beginning
of the playback process. The starvation probability
in a rebuffering process is calculated by $X_{ij}(q_b)$, given the rebuffering threshold $q_b$.

The probability of having $L$ starvations can be expressed as the product of the probabilities
from the first prefetching to the last playback. The probability vector from $I_1^A$ to $I_1^B$
is obtained by
\begin{eqnarray}
\{\mathbb{P}_{I_1^A\rightarrow I_1^B}\} =  \mathbf{\pi}\cdot \exp\big(-\theta q_a\big)\cdot\mathbf{V}(0;q_a), \; \forall
I_1^A, I_1^B \in S.
\label{eq:starvation_pgf5}
\end{eqnarray}
\noindent The probability vector from $I_1^A$ to $I_2^A$ is,
\begin{eqnarray}
&&\!\!\!\!\!\!\!\!\!\!\!\!\!\!\{\mathbb{P}_{I_1^A\rightarrow I_2^A}\} =  \{\mathbb{P}_{I_1^A\rightarrow I_1^B}\} \cdot \mathbf{X}(q_a)\nonumber\\
&& = \mathbf{\pi}\cdot \exp\big(-q_a\theta\big)\cdot\mathbf{V}(0;q_a)\cdot \mathbf{X}(q_a), \forall
I_1^A, I_2^A \in S.
\label{eq:starvation_pgf6}
\end{eqnarray}
Recall that the starvation happens at state $I_2^A$, and the rebuffering process
ends at state $I_2^B$ with the prefetched video duration $q_b$.
We next compute the probability of having only one starvation denoted by $\mathbb{P}_{\mathrm{1 starv}}$. The possible
paths include $\{I_1^A,I_1^B,I_2^A, I_e\}$ and $\{I_1^A,I_1^B,I_2^A, I_2^B,I_e\}$.
The first part of $\mathbb{P}_{\mathrm{1 starv}}$ refers to the case that the remaining video duration
is less than the rebuffering threshold $q_b$. The second part refers to the case that
the remaining video duration is longer than $q_b$ and there is no starvation after the rebuffering process.
{\small
\begin{eqnarray}
\!\!\!\!\!\!\!&&\!\!\!\!\!\mathbb{P}_{\mathrm{1 starv}} =  \{\mathbb{P}_{I_1^A\rightarrow I_2^A}\} \cdot \mathbf{1} \cdot\big(1-\exp(-q_b\theta)   \big) + \{\mathbb{P}_{I_1^A\rightarrow I_2^B}\} \cdot (1-\mathbf{W}(q_b))\nonumber\\
\!\!\!\!\!\!\!&&\!\!\!\!\!= \mathbf{\pi}\cdot \exp\big(-q_a\theta\big)\cdot\mathbf{V}(0;q_a)\cdot \mathbf{W}(q_a) \cdot\big(1-\exp(-q_b\theta)   \big)\nonumber\\
\!\!\!\!\!\!\!&&\!\!\!\!\! + \mathbf{\pi}\cdot \exp\big(-(q_a+q_b)\theta\big)\cdot\mathbf{V}(0;q_a)\cdot \mathbf{X}(q_a)\cdot \mathbf{V}(0;q_b)\cdot (1-\mathbf{W}(q_b)).
\label{eq:starvation_pgf7}
\end{eqnarray}
}
\noindent Here, the expression $(1-\mathbf{W}(q_b))$ is the probability $I_2^A\rightarrow I_e$ in the first path
and the expression $(1-\mathbf{W}(q_b))$ is that of $I_2^B\rightarrow I_e$ in the second path.
Similarly, we can deduce the probability of having $L (L>1)$ starvations recursively
by $\mathbb{P}_{\mathrm{L starv}}$
{\small
\begin{eqnarray}
\!\!\!\!\!\!\!&&\!\!\!\!\!=  \{\mathbb{P}_{I_1^A\rightarrow I_{L{+}1}^A}\} \cdot \mathbf{1} \cdot\big(1-\exp(-q_b\theta)   \big) + \{\mathbb{P}_{I_1^A\rightarrow I_{L{+}1}^B}\} \cdot (1-\mathbf{W}(q_b))\nonumber\\
\!\!\!\!\!\!\!&&\!\!\!\!\!= \mathbf{\pi}\cdot \exp\big(-q_a\theta\big)\cdot\mathbf{V}(0;q_a)\mathbf{X}(q_a)\cdot \Big(\exp\big({-}q_b\theta\big)\mathbf{V}(0;q_b)\mathbf{X}(q_b)\Big)^{L{-}1}\nonumber\\
\!\!\!\!\!\!\!&&\!\!\!\!\!\cdot \mathbf{1}\cdot \big(1-\exp(-q_b\theta)   \big) + \mathbf{\pi}\cdot \exp\big(-(q_a+q_b)\theta\big)\cdot\mathbf{V}(0;q_a)\mathbf{X}(q_a)\cdot\nonumber\\
\!\!\!\!\!\!\!&&\!\!\!\!\!\cdot \Big(\exp\big({-}q_b\theta\big)\mathbf{V}(0;q_b)\mathbf{X}(q_b)\Big)^{L{-}1}\cdot \mathbf{V}(0;q_b)\cdot (1-\mathbf{W}(q_b)).
\label{eq:finalpgf}
%
%
%
%
\end{eqnarray}
}
Though the expression in eq.\eqref{eq:finalpgf} looks complicated, it only involves duplicated
products of matrices with dimension $K$ that can be calculated easily.

\section{VBR Streaming: Modeling QoE}
\label{sec:vbr}

In this section, we investigate the QoE of variable bit rate streaming (VBR). We introduce
a diffusion process to model the variation of playback rate.

\subsection{Queueing Model of VBR Streaming}

In VBR, the frame size depends on the video scenario. For instance,
the complex segments of video clips require more bits to render each frame than
the simple segments. Then, the playback process exhibits the variation of
service rate. The complex and the simple
segments occur randomly, producing a mean playback rate. In this context,
an important question is whether the jittering of playback rate
significantly influences the starvation behavior or not.


In VBR streaming, the video file size is exponentially distributed with
the mean $1/\theta_F$. Therefore, the Markovian property
of flow departure still holds in Fig.\ref{fig:markovchain0}-\ref{fig:markovchain2}
and the transition rates remain the same as in Section \ref{sec:cbr}.
Whereas the video duration follows a general distribution.
We define the mean playback rate to be $Bitrate$. The
mean frame size is written as $\frac{Bitrate}{25}$ with frame rate 25fps. Denote by $\sigma$
the standard deviation of video frames. The total variance of video frames
is $25\sigma^2$ in one second.

We define an $It\hat{o}$ process $\{\mathcal{S}(t)\}$ to describe the total service
measured in the duration of video contents by time $t$. The $It\hat{o}$ process $\{\mathcal{S}(t)\}$ satisfies
the following stochastic differential equation
\begin{eqnarray}
d\mathcal{S}(t) = \mathcal{S}(t+h) - \mathcal{S}(t) = 1\cdot h + \bar{\sigma} d\mathcal{B}_h,
\label{eq:brownianmotion1}
\end{eqnarray}
\noindent where $\mathcal{B}$ is the standard Wiener process and the subscript $h$ denotes the duration.
The process $\mathcal{B}_h$ satisfies
$\mathcal{B}_h|_{h=0} = 0$, $E[\mathcal{B}_h] =0$ and the derivative
$d\mathcal{B}_h = \sqrt{h}\mathcal{N}(0,1)$ where $\mathcal{N}(0,1)$ is the
standard Normal distribution. 
In eq.\eqref{eq:brownianmotion1},
the parameter $\bar{\sigma}$ denotes the standard deviation of video playback
in a unit time. Hence, given the playback starting at time 0, the total variance of $\mathcal{S}(t)$
is $\mathrm{Var}[\mathcal{S}(t)] = \bar{\sigma}^2\mathrm{Var}[\mathcal{B}_t] = t\bar{\sigma}^2$.
At the unit time $t=1$ second, there has $\mathrm{Var}[\mathcal{S}(1)] = \bar{\sigma}^2$.
Remember that 25 frames are served in one second. The total variance of served bits is thus $25\sigma^2$.
When it is re-scaled by the video bitrate (measured in the duration of video contents),
the variance is expressed as $\frac{25\sigma^2}{Bitrate^2}$. Therefore, we obtain the mapping
$\bar{\sigma} = \frac{5\sigma}{Bitrate}$.

In this section, we integrate the playback perturbation with the fluid-level flow dynamics.
The method employed here is inspired by the ruin analysis in actuarial science \cite{NAA07:Lu,Dufresne}.
With the continuous time assumption, we use the diffusion process $\mathcal{S}(t) $ to
describe the queueing dynamics with the perturbation of playback rate.
The continuous time queueing process in the prefetching phase, $\{Q_a(t); t\geq 0\}$, is defined
as
\begin{eqnarray}
Q_a(t) = \sum_{l=1}^{N_e(t)} b_{I_l}(A_l - A_{l-1}) + b_{I_{N_e(t)}}(t-A_{N_e(t)}) + \bar{\sigma}\mathcal{B}_t.
\label{eq:queue_startup_diffu}
\end{eqnarray}
Similarly, the queueing process in the playback phase, $\{Q_b(t); t\geq 0\}$, is expressed as
\begin{eqnarray}
Q_b(t) = q {+} \sum_{l=1}^{N_e(t)} c_{I_l}(A_l {-} A_{l-1}) + c_{I_{N_e(t)}}(t{-}A_{N_e(t)}) {+} \bar{\sigma}\mathcal{B}_t.
\label{eq:queue_service}
\end{eqnarray}
\noindent For the VBR streaming, the starvation
can be caused by either the playback rate variation in small time scales or the flow dynamics in large
time scales.

\subsection{Starvation Probability}

The computation of starvation probability uses the similar technique as that in section \ref{sec:cbr}.
All possible events that take place in an infinitesimal time interval are taken into account.
Conditioned on the flow dynamics and throughput perturbation in $[0,h]$, we have
\begin{eqnarray}
W_i(q) \!\!&=&\!\! (1-\lambda_i h-\mu_{i} h)W_i(q+c_ih+d\mathcal{B}_h) \nonumber\\
\!\!&&\!\! + \lambda_ihW_{i+1}(q+c_ih+d\mathcal{B}_h) \nonumber \\
\!\!&&\!\!+ \nu_ihW_{i-1}(q+c_ih+d\mathcal{B}_h) + o(h), \; \forall i\in S.
\label{eq:starvprob_brownian1}
\end{eqnarray}
\noindent The above equations yield
\begin{eqnarray}
\!\!\!\!\!\!\!\!\!\!\!\!&&1/h\cdot\big(W_i(q{+}c_ih{+}d\mathcal{B}_h) {-} W_i(q)\big) = (\lambda_i{+}\mu_i) W_i(q{+}c_ih{+}d\mathcal{B}_h) \nonumber\\
\!\!\!\!\!\!\!\!\!\!\!\!&& {-} \lambda_iW_{i{+}1}(q{+}c_ih{+}d\mathcal{B}_h) {-} \nu_iW_{i{-}1}(q{+}c_ih{+}d\mathcal{B}_h) {+} o(h)/h.
\label{eq:starvprob_brownian2}
\end{eqnarray}
As $h\rightarrow 0$, the left-side of eq.\eqref{eq:starvprob_brownian2} is expressed as
\begin{eqnarray}
E[\frac{1}{h}\big(W_i(q+c_ih+d\mathcal{B}_h)- W_i(q)\big)] = c_i\dot{W}_i(q) + \frac{1}{2}\bar{\sigma}^2\ddot{W}(q),
\label{eq:starvprob_brownian3}
\end{eqnarray}
according to \cite{Dufresne}. Submitting \eqref{eq:starvprob_brownian3} to \eqref{eq:starvprob_brownian2}, we obtain
\begin{eqnarray}
&&a \ddot{W}_i(q) {+} c_i \dot{W}_i(q) {-} (\lambda_i{+}\mu_i)W_i(q){+} \lambda_iW_{i{+}1}(q) \nonumber\\
&& \;\;\;\;\;\;\;\;\;\;\;\;\;\;\;\;\;\;\;\;\;\;\;\;\;\;\;\;\;\; + \nu_iW_{i-1}(q) = 0, \forall i\in S,
\label{eq:starvprob_brownian4}
\end{eqnarray}
\noindent where $\ddot{}$ denotes the second order derivative.
The constant $a$ equals to $\frac{1}{2}\bar{\sigma}^2$.
The boundary conditions satisfy
\begin{eqnarray}
W_i(0) = 1, \;\; \forall i\in S. \\
\dot{W}_i(\infty) = 0, \;\; \forall i\in S.
\label{eq:starvprob_brownian5}
\end{eqnarray}
The starvation probability with no initial prefetching is 0 because the queueing process
is oscillating very fast. The queue length will go ``below'' 0 immediately for sure.
When $q$ is infinitely large, the starvation probability $W_i(q)$ is 0. But $W_i(q)$
approaches 0 gradually, giving rise to the first-order derivative $\dot{W}_i(\infty) = 0$. 
We denote by $\mathbf{Y}(q):=\{W_0(q),\cdots, W_{K{-}1}(q), \dot{W}_0(q),\cdots,\dot{W}_{K{-}1}(q)\}$.
We further define two matrices, $Y_3$ and $Y_4$, that have the following forms:
\begin{eqnarray}
Y_3 = \mathbf{diag}\{c_i/a\}\cdot \mathbf{M}_W \quad \textrm{and} \quad Y_4 = \mathbf{diag}\{-c_i/a\}. \nonumber
\end{eqnarray}
Then, equations in \eqref{eq:starvprob_brownian4} are rewritten in the matrix form
\begin{eqnarray}
\dot{\mathbf{Y}}(q) = \mathbf{M}_Y \mathbf{Y}(q)= \left[ \begin{array}{ccc}
\mathbf{0} & I \\
Y_3 & Y_4 \end{array} \right]\cdot \mathbf{Y}(q).
\label{eq:starvprob_brownian6}
\end{eqnarray}
The solution to eq.\eqref{eq:starvprob_brownian6} is thus given by
\begin{eqnarray}
\mathbf{Y}(q) = \exp{(\mathbf{M}_Yq)}\cdot \mathbf{Y}(0).
\label{eq:starvprob_brownian7}
\end{eqnarray}
Since $Y_3$ is similar to a symmetric tridiagonal matrix and $Y_4$ is a diagonal matrix, 
we make the following conjecture.
\begin{conj}
\label{conj:no1} The matrix $\mathbf{M}_{Y}$ has $2K$ real eigenvalues, and can be expressed
as $\mathbf{M}_{Y}=D_Y\Lambda_YD_Y^{-1}$, where $D_Y$ is an invertible matrix and $\Lambda_Y$
is a diagonal matrix.
\end{conj}
On the basis of the above conjecture, eq.\eqref{eq:starvprob_brownian7} is substituted by
\begin{eqnarray}
\mathbf{Y}(q) = D_Y\exp{(\Lambda_Yq)}D_Y^{-1}\cdot \mathbf{Y}(0).
\label{eq:starvprob_brownian8}
\end{eqnarray}

\section{Simulation}
\label{sec:simu}

In this section, we compare the numerical experiments
with the developed framework using MATLAB. Our model exhibits excellent accuracy.


\subsection{Constant bit-rate streaming}

We consider a network with maximum number of ten simultaneous streaming flows
and the capacity of 2.5Mbps.
Flows arrive to the network with a Poisson rate $\lambda = 0.12$.
Let the video duration
be exponentially distributed with the mean $60$ seconds. Then, there have
$\mu=0.1302$ and $\rho = 0.9216$ at the playback rate 360Kbps, and
$\mu=0.0868$ and $\rho = 1.3824$ at the playback rate 480Kbps.
The simulation lasts $5\times 10^5$ seconds.

\noindent\textbf{Starvation probabilities:} In this set of experiments,
we will illustrate the overall starvation probability, the
starvation probabilities when the playback process begins at
different states, as well as the p.g.f. of starvation events.

\begin{figure*}[!htb]
 \begin{minipage}{0.3\linewidth}
    \centering
   \includegraphics[width=2.4in, height = 1.9in]{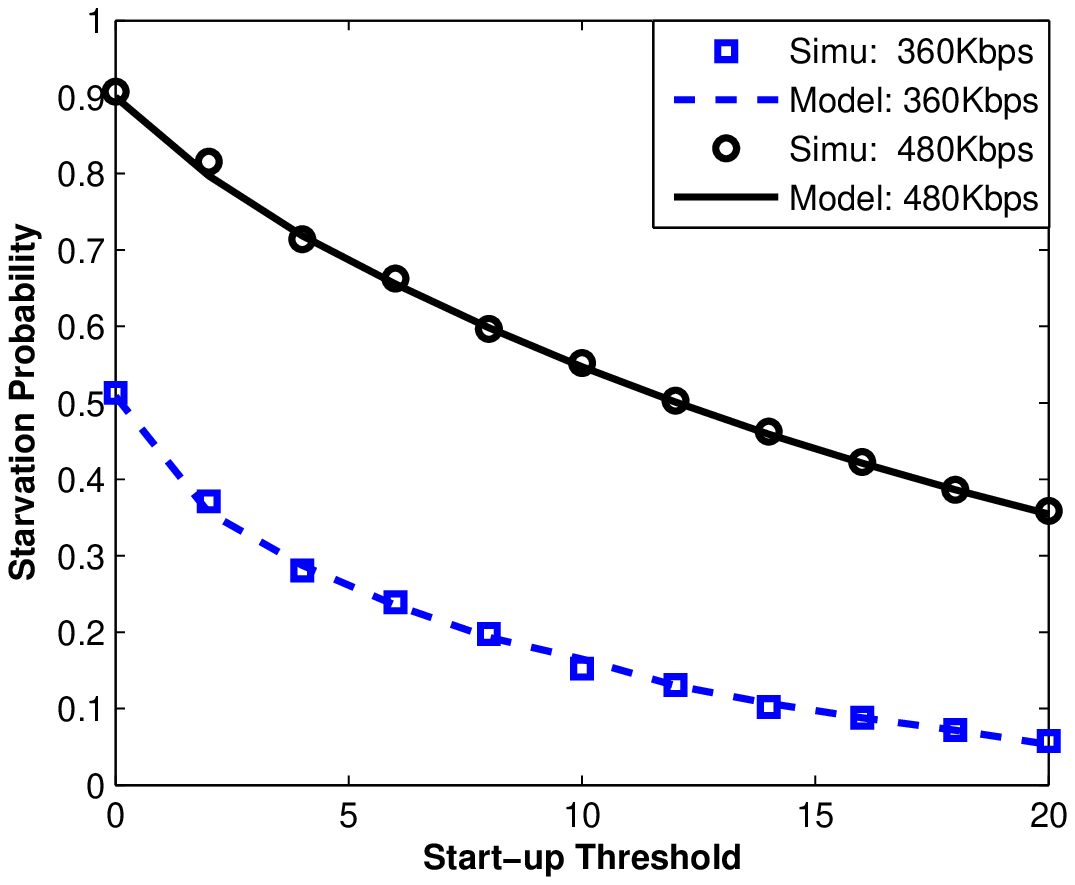}
   \caption{Overall starvation probability VS start-up threshold}
   \label{fig:starvprob1}
 \end{minipage}
 \hspace{0.5cm}
 \begin{minipage}{0.3\linewidth}
    \centering
   \includegraphics[width=2.4in, height = 1.9in]{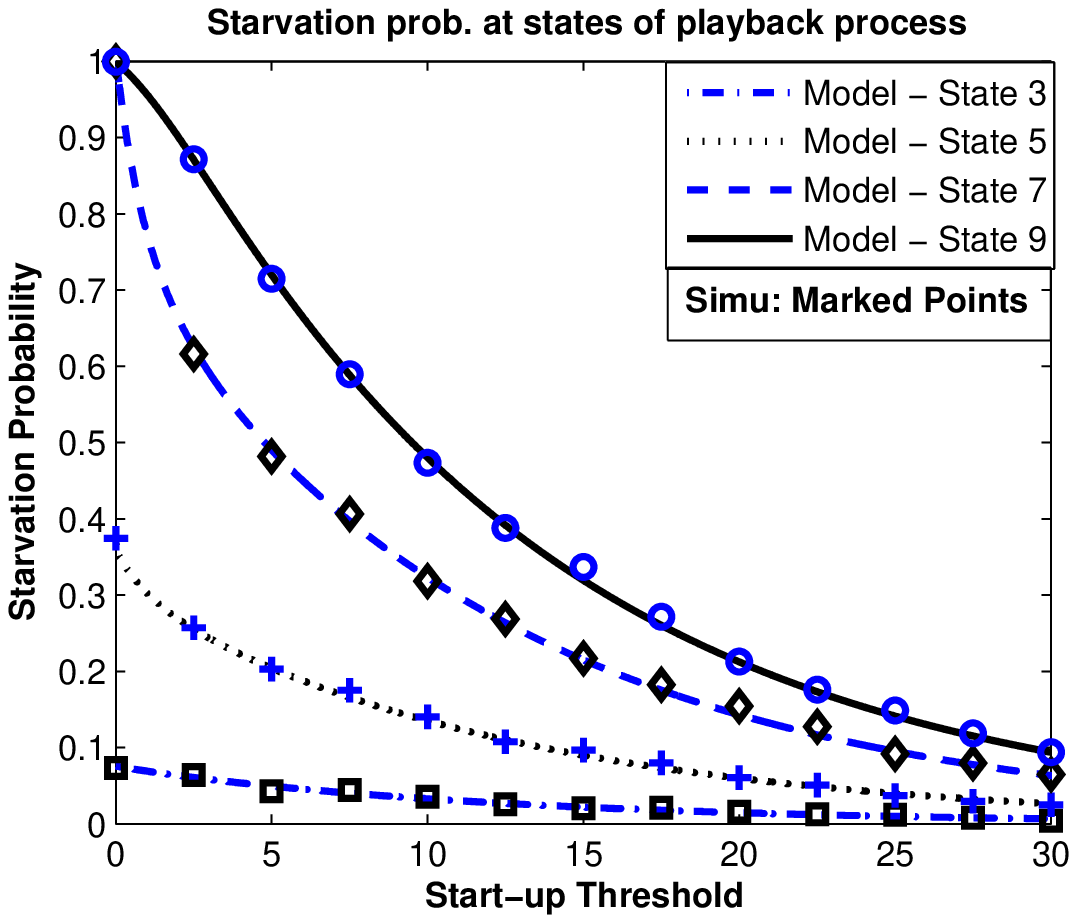}
   \caption{Starvation probabilities at different playback states with a playback rate 360Kbps}
   \label{fig:starvprob2}
 \end{minipage}
 \hspace{0.5cm}
 \begin{minipage}{0.3\linewidth}
   \centering
   \includegraphics[width=2.4in, height = 1.9in]{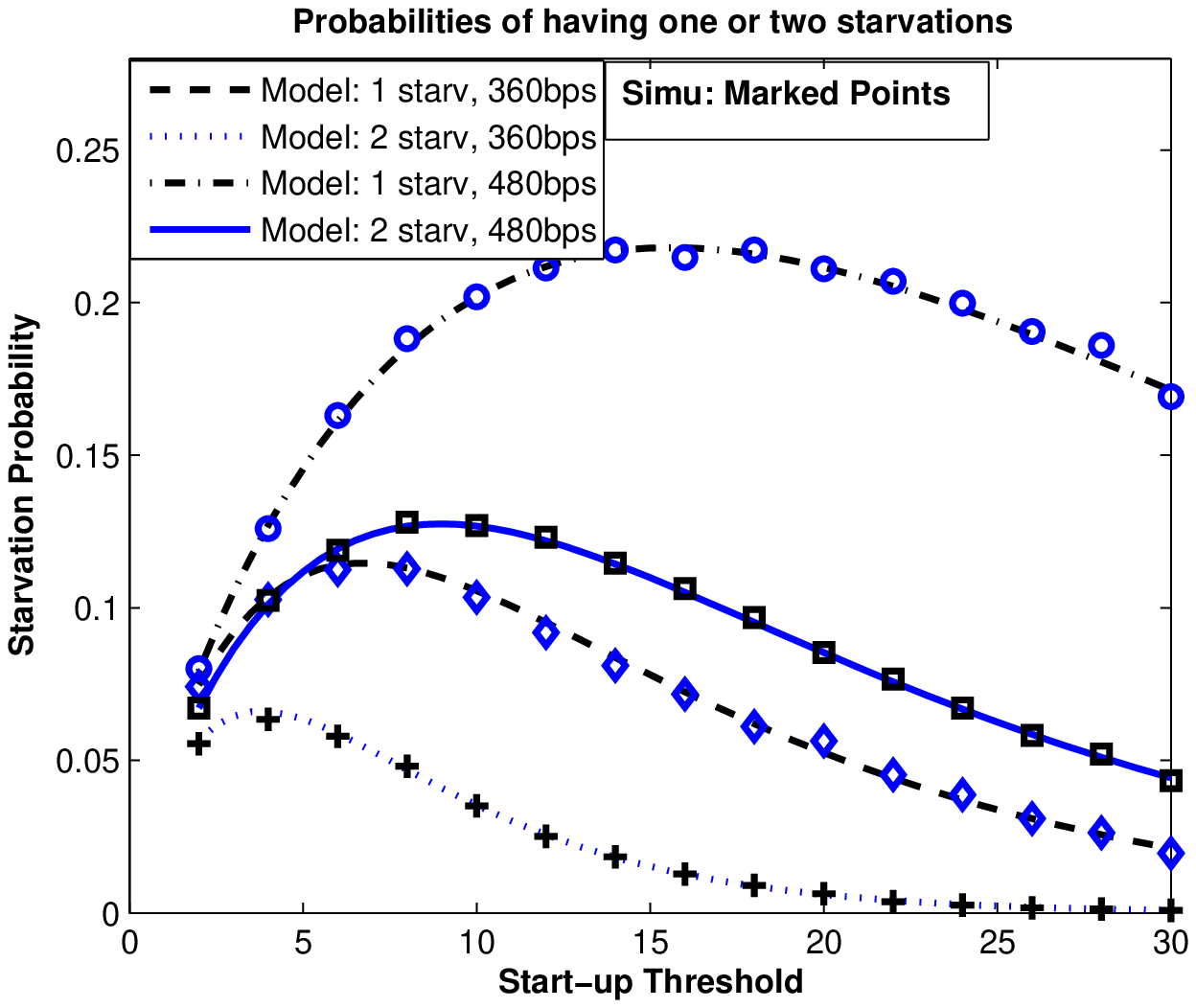}
   \caption{Probability of observing one and two starvations}
   \label{fig:starvprob3}
 \end{minipage}
\vspace{-0.3cm}
\end{figure*}

Figure \ref{fig:starvprob1} shows the overall starvation probabilities with
different settings of the start-up threshold. When it
increases from 0 to 20s of video contents, the starvation probability decreases.
The higher playback rate (e.g. 480Kbps) incurs larger starvation probabilities
in comparison with the lower playback rate (e.g. 360Kbps). Our mathematical models
match the simulations very well.

Figure \ref{fig:starvprob2} compares the starvation probabilities when
the playback process begins at different states. A higher state
refers to more coexisting flows (or congestions), and hence causing
a larger starvation probability. Note that the arrival rates at state 7 and 9 are less than 360Kbps.
Without prefetching, the starvation event happens for sure.

We further evaluate the probabilities of having one or two starvations in the whole
procedure. For clarity, we choose the same value for the start-up and re-buffering thresholds.
The starvation probabilities increase in the beginning and decrease afterwards
when $q_a$ (or $q_b$) increases from 0 to 30s of video segment. This is because there are many starvations
with very small start-up threshold and few starvations with very large start-up threshold.
Our analytical model predict the starvation probabilities accurately.

\begin{figure*}[!htb]
 \begin{minipage}{0.3\linewidth}
    \centering
   \includegraphics[width=2.4in, height = 1.9in]{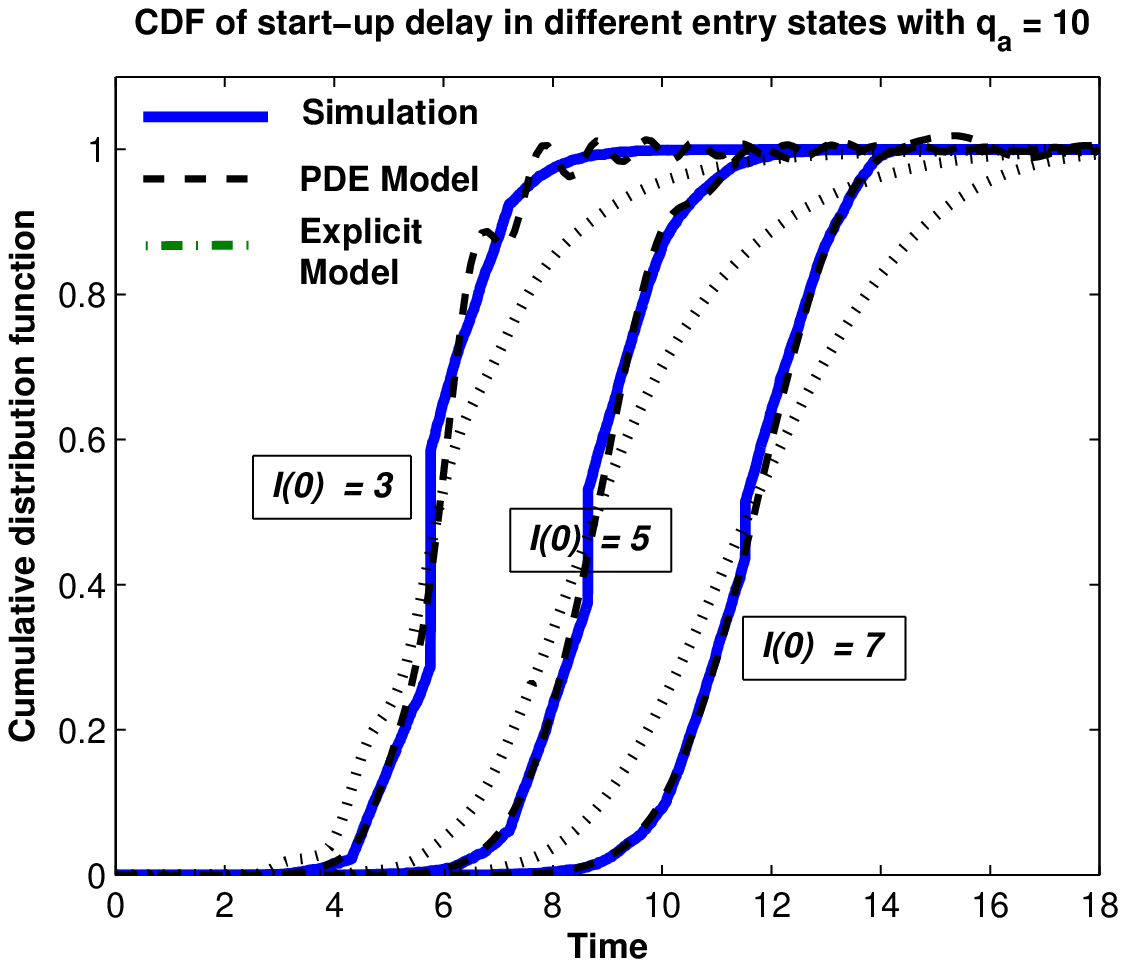}
   \caption{CDF of start-up delay with $q_a=10$}
   \label{fig:startupdelay}
 \end{minipage}
 \hspace{0.5cm}
 \begin{minipage}{0.3\linewidth}
    \centering
   \includegraphics[width=2.4in, height = 1.9in]{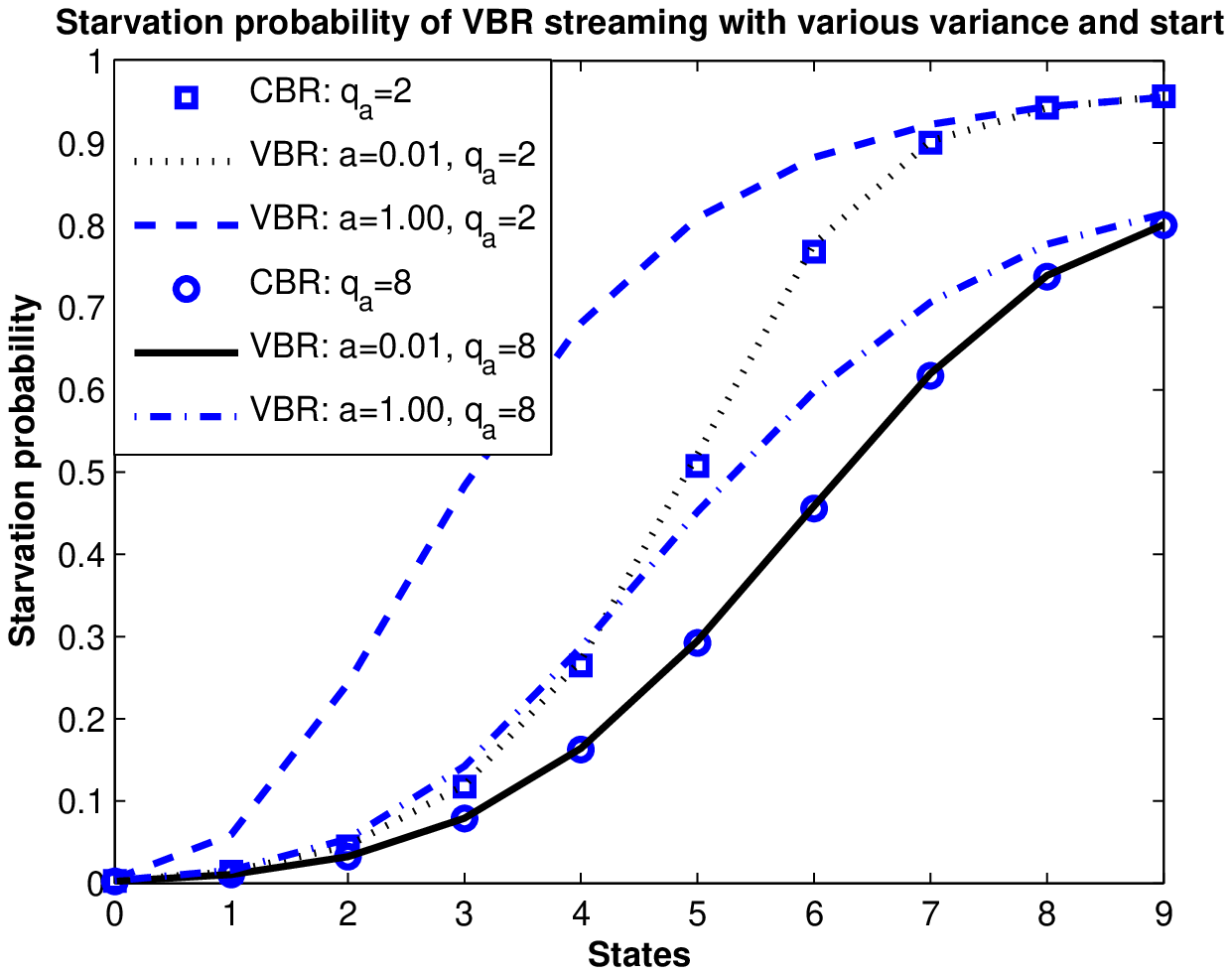}
   \caption{Starvation probabilities at all states with $d = 0.01$ and $0.5$
   computed by models}
   \label{fig:starv_vbr_1}
 \end{minipage}
 \hspace{0.5cm}
 \begin{minipage}{0.3\linewidth}
   \centering
   \includegraphics[width=2.4in, height = 1.9in]{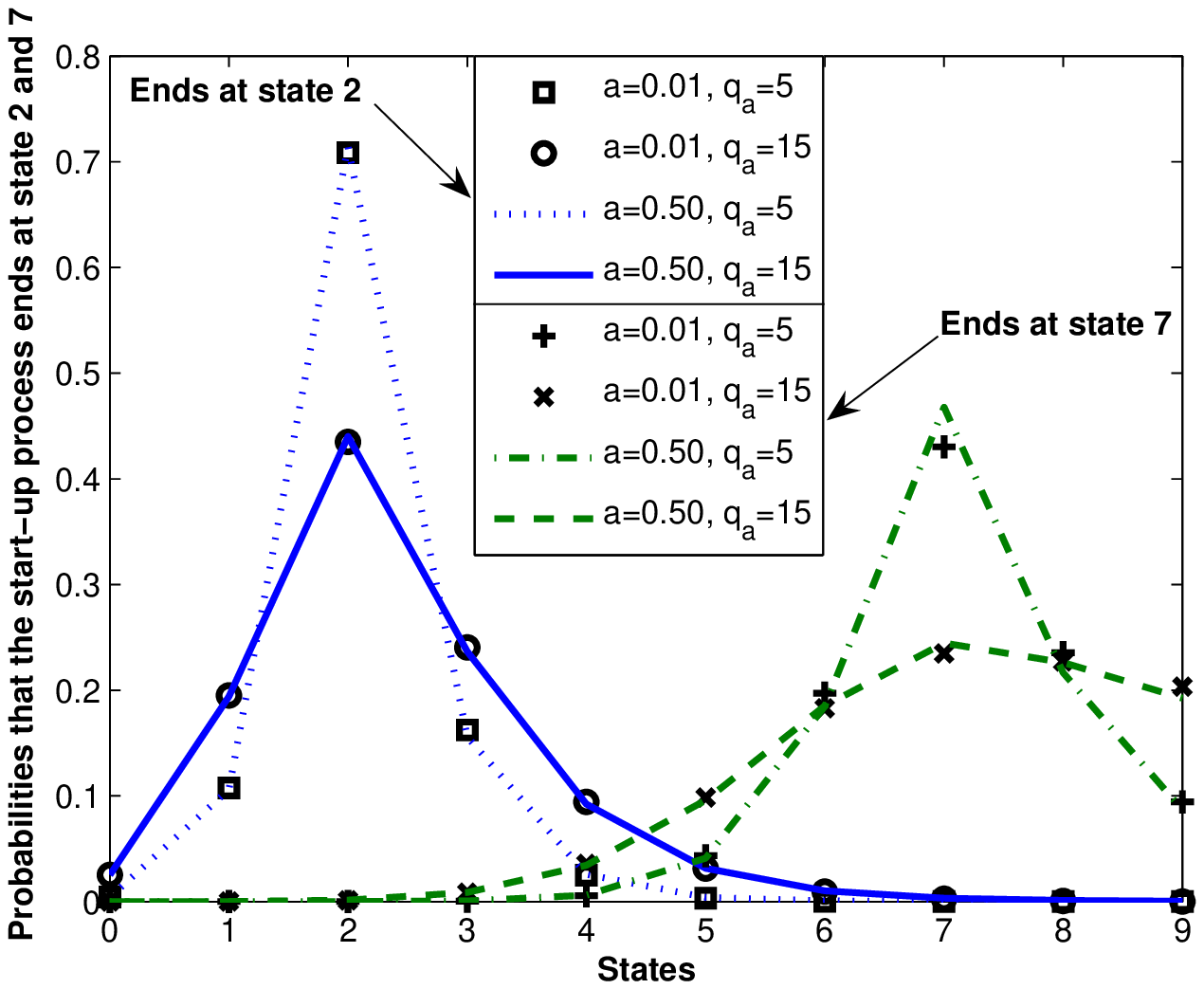}
   \caption{Probabilities that prefetching process starts from a state (from 0 to 9)
   and ends at state 2 or 7 with $d = 0.01$ and $0.5$.}
   \label{fig:startup_vbr_2}
 \end{minipage}
\vspace{-0.3cm}
\end{figure*}

\noindent\textbf{Start-up delay:} We illustrate the
distribution of start-up delays in Fig.\ref{fig:startupdelay}.
The start-up threshold is set to 10s. We highlight the
c.d.f. curves when the tagged flow sees $3,5,$ and $7$
other flows respectively after entering the network.
We use MATLAB PDE function \emph{pdepe} to compute the model in eq.\eqref{eq:solving_startupdelay_eq3}
numerically. Fig.\ref{fig:startupdelay} demonstrates accurate estimation
of start-up delay in the simulation. When the cumulative probability
is close to 1, the PDE model oscillates slightly. This is because the
initial condition $U_i(0,0)$ is discontinuous in eqs.\eqref{eq:solving_startupdelay_eq4}
and \eqref{eq:solving_startupdelay_eq6}. The dotted lines exhibit the c.d.f curves when
we adopt the Brownian motion approach to compute the explicit form. The parameter 
$\alpha$ is chosen to be $0.1$ in our paper. 
As shown in Fig.\ref{fig:startupdelay}, the explicit-form model provides a rough estimation of
the c.d.f. of start-up delay. However,  
the explicit form model has almost the same mean start-up delay as that of the experiments. 

\begin{figure*}[!htb]
 \begin{minipage}{0.3\linewidth}
    \centering
   \includegraphics[width=2.4in, height = 1.9in]{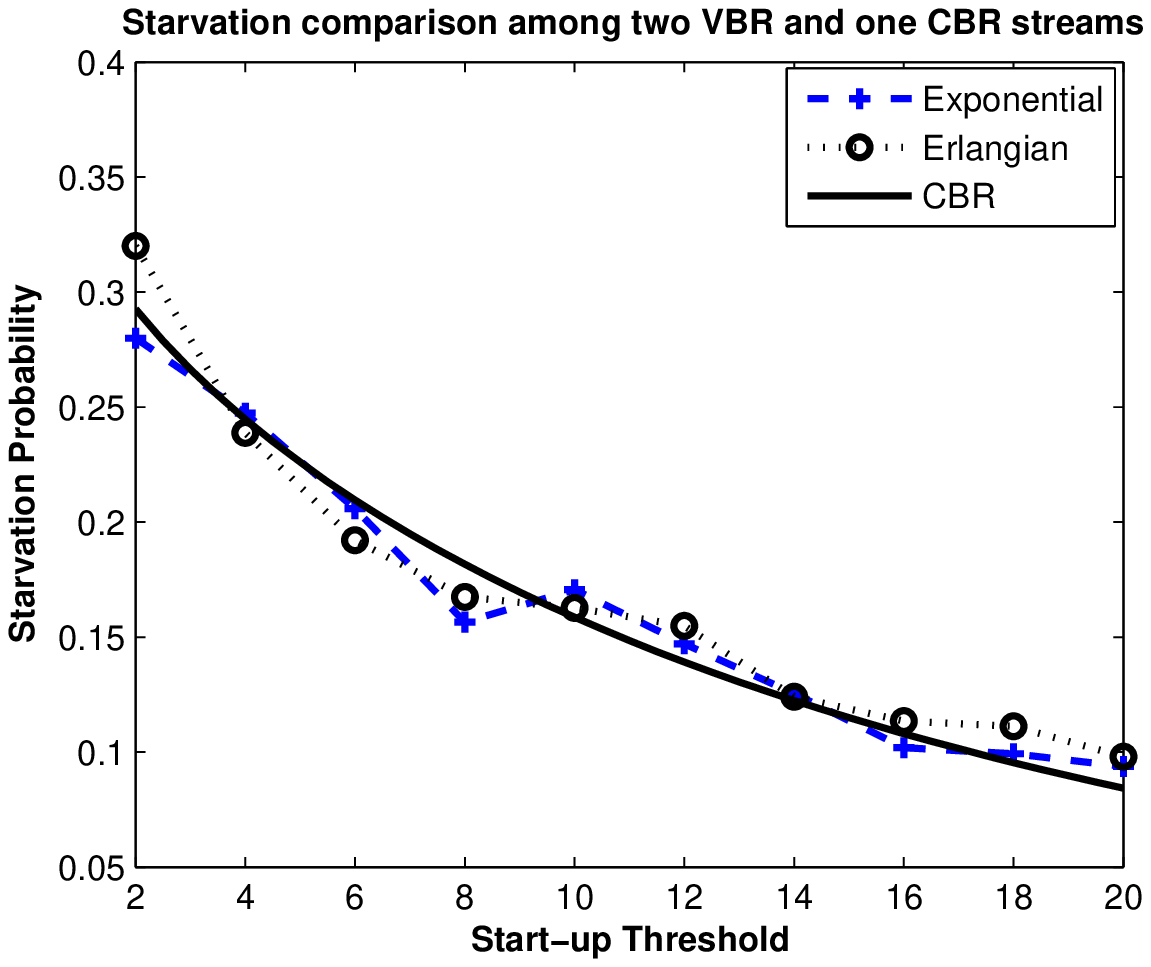}
   \caption{Starvation comparison among VBR of different frame size distributions and CBR model}
   \label{fig:starv_vbr_3}
 \end{minipage}
 \hspace{0.5cm}
 \begin{minipage}{0.3\linewidth}
    \centering
   \includegraphics[width=2.4in, height = 1.9in]{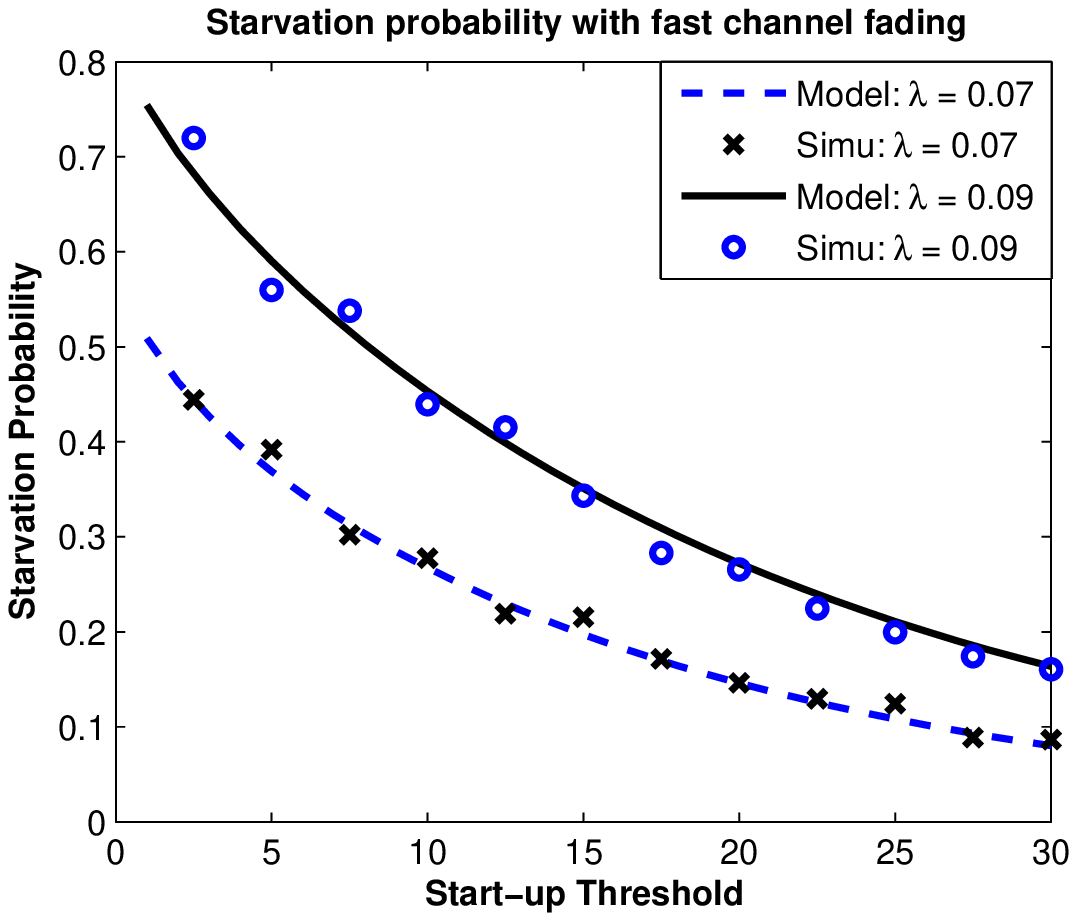}
   \caption{Starvation probability VS start-up threshold with Rayleigh fading}
   \label{fig:fastfading1}
 \end{minipage}
 \hspace{0.5cm}
 \begin{minipage}{0.3\linewidth}
   \centering
   \includegraphics[width=2.4in, height = 1.9in]{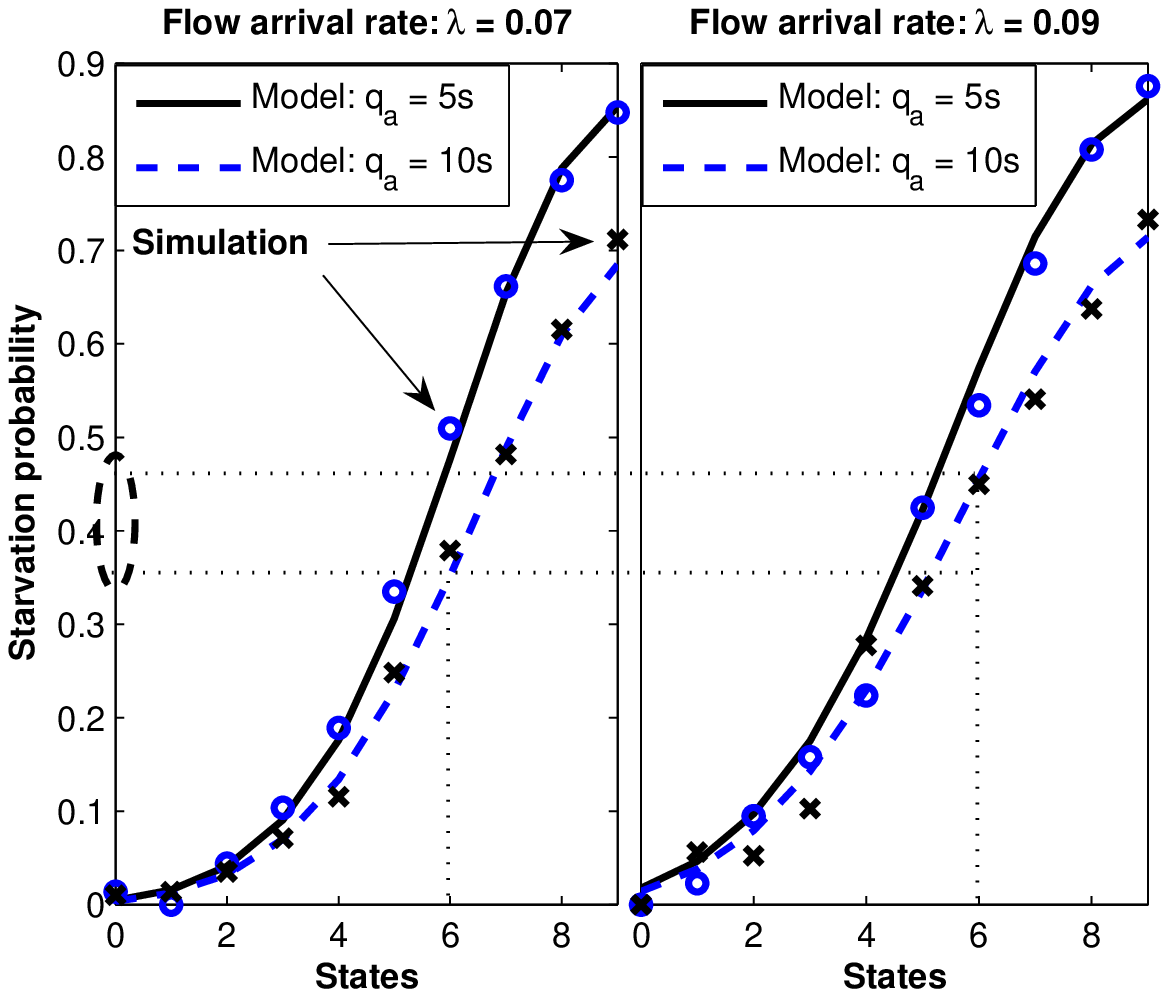}
   \caption{Starvation probability at different states with Rayleigh fading}
   \label{fig:fastfading2}
 \end{minipage}
\end{figure*}


\subsection{Variable bit-rate streaming}

We evaluate the QoE metrics of VBR streaming with a different set of parameters.
The bandwidth is set to 2.0Mbps, and the flow arrival rate is set to 0.08.
Each video streaming has the mean playabck rate of 360Kbps and a frame rate 25fps.
The size of video files are exponentially distributed with the mean $2.16\times 10^7$ bits
(equivalent to 60s with the playback rate 360Kbps). Then, the traffic load
of the system is given by $\rho = 0.864$. The per-flow throughput
in states $5\sim 9$ are insufficient to support the mean playback rate.

We first investigate how the playback variance influences
the prefetching and the playback processes. Fig.\ref{fig:starv_vbr_1}
shows the starvation probabilities when the start-up threshold
and the variance change. When $a=0.01$, the starvation probabilities
computed from the VBR model are the same as those computed from the
CBR model. While they differ greatly with $a=1$. For the case $a=1$,
the jittering of playback rate influences the starvation probability
more with $q_a=2$ than with $q_a=8$. Fig.\ref{fig:startup_vbr_2}
compares the probabilities that the prefetching process
ends at the state 2 and 7 respectively. From this set of experiment,
we can see that even $a=0.5$ does not obviously influence the prefetching.

Fig.\ref{fig:starv_vbr_3} compares the numerical results of VBR streaming
with the model for CBR streaming.
In our simulation, the mean frame size is 14400 bits. According to \cite{MASI08},
the video frame size roughly follows Erlang distribution.
If the Erlang distribution is the sum of $k$ i.i.d. exponentially distributed r.v.s.,
the mean of these r.v.s. is $14400/k$. We consider two cases in this set of
experiments, $k=1$ (i.e. exponential r.v.) and $k=3$. The resulting variances
are $\bar{\sigma}^2 = 0.04$ (i.e. $a = 0.02$) for $k=1$ and $\bar{\sigma}^2 = 0.013$
(i.e. $a=0.0066$) for $k=3$.
The simulation time is $3\times 10^6$ playback slots. From Fig.\ref{fig:starv_vbr_3},
we are surprised to see
that the Erlang distributions of video frames do not obviously influence
the starvation probabilities. The analytical framework for CBR streaming is
good enough to model the starvation behavior for VBR streaming.

\section{Extension to Fast Fading}
\label{sec:extension}

This section models the starvation behavior of CBR streaming
when users experience fast channel fading.
We compute the first two moments of bit arrival process and
show how these parameters can be feed into our analytical
framework.

{\bf Network description.}
Due to the change of radio condition (e.g. user mobility, or a car passing by the user),
the signal strength is no longer a constant at different scheduling slots. To explore
the multiuser diversity gain, the base station adopts the normalized SNR scheduling algorithm
for allocating time slots to coexisting flows.

We begin with the scenario with a fixed population of $i$ users (or flows)
served by a single base station. In each slot, the users measure their channel
qualities and feedback them to the BS. Based on the channel quality indications,
the BS transmits to only one of the users every slot.
Denote by $\gamma_{j,n}$ the instantaneous signal
to noise ratio (SNR) of user $j$, ($1\leq j\leq i$), at slot $n$.
As stated in most of previous work,
we assume that all the users experience Rayleigh fast-fading.
Denote by $\bar{\gamma}_j$ the average SNR of user $j$. Then, the received SNR
of user $j$ is an exponentially distributed random variable with the
following probability density function
$ g_j(\gamma) = \frac{1}{\bar{\gamma}_j}\exp(-\frac{\gamma}{\bar{\gamma}_j}).
$
The NSNR scheduler selects the user that has the highest relative SNR for transmission,
$
j^*_n = \max_j\{\gamma_{j,n}/\bar{\gamma}_j,\; j=1,2,\cdots,i\},
$
where $j^*$ is the scheduled user at slot $n$. In this section, we consider
the case of homogeneous average SNRs (i.e. $\bar{\gamma}_j=\bar{\gamma}$ for all $j$). Therefore, the NSNR scheduler is equivalent to
the maximum sum rate (MSR) scheduler that gives the largest per-user throughput.
Since the SNRs of different users are independently distributed, the scheduled SNR,
denoted by $\gamma^*$, has the following probability density function \cite{JSAC07:Chang}
$ g^*(\gamma) = \frac{i}{\bar{\gamma}}\exp(-\frac{\gamma}{\bar{\gamma}}) \big(1-\exp(-\frac{\gamma}{\bar{\gamma}})\big)^{i{-}1}. $
Denote by $f(\gamma)$ the data rate of a user with the SNR $\gamma$. Here, $f(\cdot)$ can be a linear function in the low-SNR regime
and a logarithmic function in the high SNR regime if the modulation scheme is continuous. For discrete modulations,
$f(\cdot)$ is a step function of $\gamma$. Without loss of generality, we let $f(\gamma) = \log_2(1+\gamma)$. 

{\bf Analysis of throughput process.}
The fast fading along with NSNR scheduling brings variation of bit arrivals to the receiver.
The analytical framework for VBR streaming can be naturally extended to
this scenario. The only modification lies in that the jittering of playback rate
is substituted by that of bit arrivals. Therefore, we need the knowledge
of the mean throughput and its variance measured
in the duration of video contents. To achieve this goal, we must obtain
the mean throughput and its variance measured in bits first.

Denote by $r_i^*$ the transmission rate of the user with the best SNR at a slot
in each Hz when
there are $i$ active flows in the cell. Denote by $r_{i}$ the transmission rate
to \underline{\emph{one particular flow}} at a slot per Hz. Given the assumption that all the flows have the same
average SNR, each flow has the equal probability of being scheduled. Hence, we can see
\begin{eqnarray}
r_{i} := \left\{\begin{matrix}
\;r_i^* \;\; &&\textrm{ w.p. } \;\;\frac{1}{i} ;\\
\;0 \;\; &&\;\textrm{ w.p. } \;\;\frac{i-1}{i}.
\end{matrix}\right.
\label{eq:pf_thru1}
\end{eqnarray}
For the r.v. $r_i^*$, its mean and variance are computed by
\begin{eqnarray}
E[r_i^*] \!&=&\! \int_0^{\infty} f(\gamma)\cdot g^*(\gamma) d\gamma, \label{eq:pf_thru2}
\end{eqnarray}
\begin{eqnarray}
\mathrm{Var}[r_i^*] \!&=&\! \int_0^{\infty} f(\gamma)^2\cdot g^*(\gamma) d\gamma - (E[r_i^*])^2.
\label{eq:pf_thru3}
\end{eqnarray}
The eqs \eqref{eq:pf_thru1}-\eqref{eq:pf_thru3} yield
\begin{eqnarray}
E[r_i] \!&=&\! \frac{1}{i} E[r_i^*], \label{eq:pf_thru4}\\
\mathrm{Var}[r_i] \!&=&\! E[r_i^2] - (E[r_i])^2 = \frac{1}{i}E[(r_i^*)^2]-\frac{1}{i^2}(E[r_i^*])^2 \nonumber\\
\!&=&\!  \frac{1}{i}\mathrm{Var}[r_i^*] + (E[r_i^*])^2(\frac{1}{i}-\frac{1}{i^2}).
\label{eq:pf_thru5}
\end{eqnarray}
Denote by $D_s$ the duration of scheduling slot (usually 2ms), and by $B$ the width of wireless spectrum
in Hz. Then, the mean and the variance of per-flow throughput measured in the \emph{duration of video contents} are
$\frac{B\cdot D_s\cdot E[r_i]}{Bitrate}$ and $(\frac{B\cdot D_s}{Bitrate})^2\cdot \mathrm{Var}[r_i]$
respectively in one slot.

Let $R_i$ be the r.v. of \underline{\emph{per-flow throughput in one second}} that is measured by the duration of
video contents. In one second, the total throughput of a flow at one Hz is the sum of throughput in $\frac{1}{D_s}$ slots.
Therefore, the r.v. $R_i$ is the sum of $\frac{1}{D_s}$ i.i.d. r.v.s corresponding to the per-slot throughput.
We can express the mean and the variance of $R_i$ as follows:
\begin{eqnarray}
\!\!\!E[R_i] \!\!\!\!&=&\!\!  \frac{1}{D_s}\cdot \frac{B\cdot D_s\cdot E[r_i]}{Bitrate} = \frac{B\cdot E[r_i^*]}{i\cdot Bitrate}, \label{eq:pf_thru6}\\
\mathrm{Var}[R_i] \!\!\!\!&=&\!\! \frac{1}{D_s}\cdot (\frac{B\cdot D_s}{Bitrate})^2\cdot \mathrm{Var}[r_i] \nonumber\\
\!\!\!\!\!\!\!\!&=&\!\!\!\! \big(\frac{1}{i}\mathrm{Var}[r_i^*] {+} (E[r_i^*])^2(\frac{1}{i}{-}\frac{1}{i^2})\big)\cdot\frac{B^2\cdot D_s}{Bitrate^2}.
\label{eq:pf_thru7}
\end{eqnarray}
In general, the frequency width $B$ is 1$\sim$5 MHz, the bit-rate is usually greater than
200 Kbps, and $D_s$ equals to 0.002s. Then, $\mathrm{Var}[R_i]$ is usually at the order of $10^{-2}$.
If starvation happens at state $i$,
$E[R_i]$ is usually less than 1, which means that $\frac{B}{Bitrate}$ needs to be small.
However, the small $\frac{B}{Bitrate}$ results in the small variance $\mathrm{Var}[R_i]$.
This is to say, if the variance of bit arrival process is large, there might not exist starvations.
On the contrary, if the starvations appear, the variance is usually small so that its impact
on the starvation is negligible. For this reason, we directly use the framework without diffusion approximation
to model the streaming QoE in a fast fading channel.

{\bf Markov model of flow dynamics}
To analyze the interaction between NSNR scheduling and the flow dynamics,
a fluid-level capacity model is required. When the average SNR
of all active users are the same, the per-flow throughput in each slot
is i.i.d. and only depends on the quantity of flows (see eq.\eqref{eq:pf_thru2}).
Given the exponentially distributed video size, we can model the flow dynamics
as a Markov process.

The Markov processes in Fig.\ref{fig:markovchain0}-\ref{fig:markovchain2}
contain transitions rates such as $\mu_i,\nu_i$ and $\varphi_i$. However,
it is not direct to feed the parameters of this section into the above Markov processes.
In Fig.\ref{fig:markovchain0}, state $i$ refers to the number of flows in the system.
The departure rate is computed by $\mu_i = i\theta E[R_i]$ for $i\in S\cup\{K\}$,
recalling that $E[R_i]$ is average per-user throughput in video duration per second.
It is easy to obtain the stationary distribution of having $i$ flows by
{\small
\begin{eqnarray}
z_i^a \!\!&=&\!\! \frac{\lambda^i}{\prod_{l=1}^{i}\mu_l}
\left[ 1+\sum_{j=1}^{K}\frac{\lambda^j}{\prod_{l=1}^{j}\mu_l}
\right]^{-1} , \;\;\; \forall i=0,\cdots,K,\nonumber
\label{eq:pf_stationary1}
\end{eqnarray}
}
(with the convention that $\prod$ over an empty set is 1).
When a tagged user joins in the system and is also admitted, it observes $i$
other flows with the following stationary distribution $\{\pi\}:$
\begin{eqnarray}
\pi_i = \frac{z_i^a}{1-z_K^a} = \frac{\frac{\lambda^i}{\prod_{l=1}^{i}\mu_l}}{1+\sum_{j=1}^{K-1}\frac{\lambda^j}{\prod_{l=1}^{j}\mu_l}},\;\; \forall i\in S. \nonumber
\label{eq:pf_stationary2}
\end{eqnarray}

The Markov processes shown in Fig.\ref{fig:markovchain1}-\ref{fig:markovchain2}
are conditioned on the existence of the tagged flow. At state $i$, the per-user
throughput is $E[R_{i+1}]$ because there are $i$ flows plus the tagged one.
Hence, the transition rate $\nu_i$ is computed by $\nu_i := i\theta\cdot E[R_{i+1}]$
for all $i\in S$. The transition rate $\varphi_i$ is expressed as $\varphi_i:=\theta\cdot E[R_{i+1}]$.
Define $\tilde{\mu}_i$ as the total departure rate at state $i$ that has
\begin{eqnarray}
\tilde{\mu}_i:=\varphi_i+\nu_i = (i+1)\theta E[R_{i+1}] = \mu_{i+1},
\label{eq:pf_transitionrate1}
\end{eqnarray}
in the presence of the tagged flow. The constants $b_i$ and $c_i$ are obtained by
\begin{eqnarray}
b_i = E[R_{i+1}] \;\; \textrm{ and } \;\; c_i = b_i - 1,\;\; \forall i\in S.
\label{eq:pf_transitionrate2}
\end{eqnarray}
Substituting the above parameters to the framework in section \ref{sec:cbr},
we can derive the approximated QoE metrics in a fast fading channel with flow dynamics.

{\bf Numerical Examples.}
Consider a wireless channel with frequency width of 1MHz.
The average SNRs of users is 5dB. The base station allows at most 10 flows simultaneously,
and schedules the transmission to one of them in every slot of duration 0.002s.
The video duration is exponentially distributed with the mean of 90 seconds and the
video bit rate is chosen to be 480Kbps. Then, the mean throughput
are $\{$3.5749, 2.3702, 1.7844, 1.4369, 1.2061, 1.0412, 0.9174, 0.8207, 0.7432, 0.6794$\}$
times the playback rate at states from 0 to 9. In other words, the mean throughput
at states 6$\sim$9 are insufficient to support the continuous playback.
The variances at all states are
$\{$0.0083, 0.0144, 0.0144, 0.0134, 0.0124, 0.0114, 0.0105, 0.0098, 0.0091, 0.0086$\}$,
which are small enough. We consider two flow arrival rates, $\lambda = 0.07$ and $\lambda = 0.09$.
For $\lambda = 0.07$, the traffic load $\rho$ is greater than 1 at states 0$\sim$5
and less than 1 at states 6$\sim$9. For the latter case, there have $\rho>1$ at all the states.
Each set of simulation lasts $2\times 10^7$ time slots.

In Fig.\ref{fig:fastfading1} we compare the starvation probabilities measured
from a Rayleigh fading channel, and those computed from the model without
considering throughput variation. The simulation matches the model quite well,
which means that the flow-level dynamics have a dominant impact on the playback interruption,
while the impact of throughput variation due to Rayleigh fading is negligible.
In Fig.\ref{fig:fastfading2} we examine the starvation probabilities when the playback process
begins at different states. We test two start-up thresholds, $q_a =\{5, 10\}$, and two flow arrival
rates, $\lambda=\{0.07, 0.09\}$. One can observe
that the starvation probabilities do not differ much in high states (e.g. 8 and 9).
However, the starvation probabilities in the states with mean throughput around 1
are distinguishable, in which state 6 is an example.
With $\lambda=0.09$, a tagged flow sees the congested network (more other flows)
with a higher probability, and also encounters a higher probability of starvation afterwards.

\section{Conclusions and Further Extensions}
\label{sec:conclusion}

In this work, we developed an analytical framework to compute the
QoE metrics of media streaming service in wireless data networks.
Our framework takes into account the dynamics of playout buffer at
three time scales, the scheduling duration, the
video playback variation, as well as the flow arrivals or departures.
We show that the proposed models can accurately predict the distribution of prefetching delay
and the probability generating function of buffer starvations.
The analytical results demonstrate that the flow dynamics have dominant influence
on QoE metrics compared to the jittering in the throughput and the video playback rate.

\textbf{Further Extensions:} Our analytical framework can be adapted to the following scenarios: i) hyper-exponential video length distribution,
ii) heterogeneous channel gains, and iii) mixed data and
streaming flows. The heterogeneity of video durations, channel gains, and traffic types
requires the classification of flows. The heterogeneous
video duration is usually modeled by the hyper-exponential
distribution. Users requesting the videos of the same exponential distribution fall
in one class. The same argument holds in the case of heterogeneous SNRs among users.
We can group the users with more or less the same average SNR in the same class (e.g. see \cite{Borst}).
The service times are still exponentially distributed, but with different parameters
in different user classes.
When classes are introduced, the Markov process are thus
modified to contain multi-dimensional states,
representing the number of (observed) flows in different classes.
We can then construct the PDEs and the ODEs on top of them.



\bibliographystyle{abbrv}

\section*{Appendix}

\subsection{Solving PDEs}

Suppose that $U_i(q,t)$ is a function of variable $r$ where $q$ and $t$ are expressed as 
$q(r)$ and $t(r)$. We take first-order derivative of $U_i(q,t)$ over $r$ and obtain
\begin{eqnarray}
\frac{dU_i}{dr} = \frac{\partial U_i}{\partial q} \frac{dq}{dr}  + \frac{\partial U_i}{\partial t}\frac{dt}{dr}.
\label{eq:pdes_appendix1}
\end{eqnarray}
We first solve the following homogeneous PDEs originated from Eq.\eqref{eq:solving_startupdelay_eq3}
\begin{eqnarray}
\frac{\partial U_i}{\partial t} + b_i \frac{\partial U_i}{\partial q} =0, \;\;\; \forall i\in S,
\label{eq:pdes_appendix2}
\end{eqnarray}
Comparing Eqs.\eqref{eq:pdes_appendix1} with Eqs.\eqref{eq:pdes_appendix2}, we have
\begin{eqnarray}
\frac{dq}{dt} = b_i \quad \textrm{and} \quad \frac{dt}{dr} = 1.
\label{eq:pdes_appendix3}
\end{eqnarray}
The above simple differential equations give rise to
\begin{eqnarray}
t = t_0 + r \quad \textrm{and} \quad  q = q_0^{(i)} + b_i r. \nonumber
\label{eq:pdes_appendix4}
\end{eqnarray}
In general, $t_0$ is set to 0 such that there have
\begin{eqnarray}
t = r \quad \textrm{and} \quad  q = q_0^{(i)} + b_i t. \nonumber
\label{eq:pdes_appendix4}
\end{eqnarray}
Then, $U_i$ is a function of the variable $q_0^{(i)} $. Define $F_i(\cdot)$ to be 
a continuous and differentiable function in the range $[-\infty, +\infty]$. $U_i$
is solved by 
\begin{eqnarray}
U_i(q,t) = F_i(q_0^{(i)}) = F_i(q-b_it)
\label{eq:pdes_appendix5}
\end{eqnarray}
when the PDEs are homogeneous.
We next proceed to consider the inhomogeneous parts at Eqs.\eqref{eq:solving_startupdelay_eq3} 
in the matrix form
\begin{eqnarray}
d\mathbf{U}/dr = -\mathbf{M}_S.
\label{eq:pdes_appendix6}
\end{eqnarray}
Then, there has 
\begin{eqnarray}
\mathbf{U}(q,t) \!\!\!&=&\!\!\! \exp(-\mathbf{M}_Sr)\cdot U_0 = \exp(-\mathbf{M}_St)\cdot U_0 \nonumber\\
\!\!\!&=&\!\!\! \exp(-\mathbf{M}_St)\cdot \{F_i(q-b_it)\}
\label{eq:pdes_appendix7}
\end{eqnarray}
where $\{F_i(q-b_it)\}$ denotes a column vector of $F_i(q-b_it)$ for all $i=0,\cdots,K{-}1$. 
Due to the discontinuity of $U_i(q,t)$ at the point $(q,t)=(0,0)$, we propose to solve it using the known 
results in Brownian motion. The arrival rate of streaming packets at state $i$ is $b_i$, measured in seconds. 
We add a very small variance to the arrival rate where the standard deviation is denoted by $\alpha$. We 
use Brownian motion to approximate the arrival rate of streaming
packets. According to \cite{TMM10:Luan,CZ}, the solutions of the homogeneous PDEs are approximated by
\begin{eqnarray}
F_i(x) = \Phi(x) = \frac{1}{\sqrt{2\pi}}\int_{-\infty}^{x}e^{-y^2/2}dy = \frac{1}{2}\textbf{erfc}(-\frac{x}{\sqrt{2}}).
\label{eq:pdes_appendix8}
\end{eqnarray}
given $\alpha$ is small enough.
Submitting Eq.\eqref{eq:pdes_appendix8} to Eq.\eqref{eq:pdes_appendix7}, we solve the inhomegeneous PDEs by
\begin{eqnarray}
\mathbf{U}(q,t) \!\!&=&\!\! \exp{(-\mathbf{M}_St)}\cdot \{\mathbf{F}_i(-\frac{q}{b_i}+t)\} \nonumber\\
\!\!&=&\!\! D_S\exp{(-\Lambda_St)}D_S^{-1}\cdot \{\mathbf{F}_i(-\frac{q}{b_i}+t)\} \nonumber\\
\!\!&=&\!\! \frac{1}{2}D_S\exp{(-\Lambda_St)}D_S^{-1}\cdot \{\textbf{erfc}(-\frac{q-b_it}{\sqrt{\alpha t}})\}.
\label{eq:pdes_appendix11}
\end{eqnarray}

\subsection{Proof of Lemma 1}

\noindent \textbf{Proof:}  
Without loss of generality, we consider a tridiagonal matrix $T$ in the form
\begin{eqnarray}
T = \left( \begin{array}{cccccc}
x_1 & y_1 & 0 & \cdots & 0 & 0 \\
z_2 & x_2 & y_2 & \cdots & 0 & 0 \\
\cdots & \cdots & \cdots & \cdots & \cdots & \cdots \\
0 & 0 & \cdots & \cdots & z_N & x_N \end{array} \right) \nonumber
\end{eqnarray}
where $x_i,y_i,z_i$ are all real constants. 
Our claim is a natural conclusion of the following lemma. 
\begin{lemma}
\label{lemma:general_tridiagonal}\cite{Geist}
Assume that the coefficients $y_i$, $i=1,\cdots,N-1$ are nonzero, and
the products $y_iz_i$ are positive. Then, the matrix $T$ is similar to a
symmetric tridiagonal matrix. Therefore, its eigenvalues are all real.
\end{lemma}
Here, $\mathbf{M}_V$ satisfies the conditions in the above lemma. Thus,
$\mathbf{M}_V$ is similar to a symmetric matrix, and is diagonizable. 
According to Gershgorin circle theorem \cite{matrixbook},
every eigenvalue of $\mathbf{M}_V$ lies within at least one of the Gershgorin discs. 
Because the diagonal element is positive, and is larger than the sum of absolution 
values of non-diagonal elements in each line, every eigenvalue cannot be 
negative. This concludes the proof. \done

\end{document}